\documentclass[12pt,preprint]{aastex}
\title{Hierarchical Structure Formation and Chemical Evolution of Galaxies}
\author{Y.-Z. Qian\altaffilmark{1} and G. J. Wasserburg\altaffilmark{2}}
\altaffiltext{1}{School of Physics and Astronomy, University of
Minnesota, Minneapolis, MN 55455; qian@physics.umn.edu.}
\altaffiltext{2}{The Lunatic Asylum, Division of Geological and
Planetary Sciences, California Institute of Technology, Pasadena,
CA 91125.}

\received{2004 February 24}
\begin{document}

\begin{abstract}
We present an analytical and phenomenological model for metal
enrichment in halos based on hierarchical structure formation.
This model assumes that astration of normal stellar populations
along with Type II supernovae (SNe II) already occurs at very high
redshift. It focuses on the regime of [Fe/H]~$<-1$ where SNe Ia
are not major contributors. The regime of [Fe/H]~$>-1$ where SNe Ia
are major contributors is also discussed in general.
For halos that are not disrupted by SN II 
explosions, the chemical evolution of the gas and stars is explicitly 
determined by the rate of gas infall as compared with the astration 
rate and the corresponding rate of metal production by SNe II per H atom 
in the gas. This model provides a good description of the data on [Fe/H]
for damped Ly$\alpha$ systems over a wide range of redshift $0.5<z<5$.
For all halos not disrupted by SN II explosions, if there is a 
cessation of gas infall, the metallicities of stars follow a bimodal 
distribution. This distribution is characterized by a sharp peak at the 
value of [Fe/H] corresponding to the time of infall cessation and by a
broad peak at a higher value of [Fe/H] corresponding to the
subsequent period of astration during which the bulk of the
remaining gas forms stars. Such a distribution may be compared to that 
observed for the Galactic halo stars.
If the gas in a halo is rapidly lost upon cessation
of infall, then an assemblage of stars with a very sharply-defined [Fe/H] 
value will be left behind. This assemblage of stars may be accreted by a 
larger system and become a globular cluster of the larger system. We also 
discuss the masses and metallicities of the globular clusters in this model 
considering the criterion for the onset of astration and possible disruption 
of low-mass halos by SN II explosions. The implications of possible globular 
clusters with very low metallicities of [Fe/H]~$\lesssim -3$ are discussed.
\end{abstract}


\section{Introduction}
We present an analytical and phenomenological model for chemical evolution
of galaxies based on hierarchical structure formation. This model provides 
a possible explanation for the metal abundances observed in damped 
Ly$\alpha$ (DLA) systems and the metallicity distribution of stars in
galaxies. It also suggests a possible mechanism for the formation of 
globular clusters. Measurement of ``metallicities'' of DLA systems by 
Lu et al. (1996) and further studies by Prochaska \& Wolfe (2000, 2002)
showed that these systems have a large dispersion in 
[Fe/H]~$\equiv\log{\rm (Fe/H)}-\log{\rm (Fe/H)}_\odot$ for a given 
redshift $z$ and that there appears to be a baseline enrichment at the 
level of [Fe/H]~$\sim -3$ over a wide range of redshift
$1.5\lesssim z\lesssim 4.5$. The work by Prochaska et al. (2003) provided 
a much more extensive data base over a wider range of redshift $0.5<z<5$.
All these data exhibited the same behavior. We earlier found that
the data on DLA systems could be explained by a simple 
closed-system model (Wasserburg \& Qian 2000b and Fig. 1 here), 
where the number abundance 
(Fe/H) of Fe relative to H atoms in a DLA system is determined by 
\begin{equation}
{\rm (Fe/H)}={\rm (Fe/H)}_P+\lambda_{\rm Fe}[t(z)-t^*].
\label{cdla}
\end{equation}
In the above equation, (Fe/H)$_P\sim 10^{-3}{\rm (Fe/H)}_\odot$ 
([Fe/H]$_P\sim -3$) is the initial Fe abundance corresponding to a 
postulated prompt inventory of metals in the intergalactic medium (IGM),
$\lambda_{\rm Fe}$ is the rate of Fe production by Type II supernovae
(SNe II) per H atom in the gas of the DLA system and is taken to be 
$\sim (30\ {\rm Gyr})^{-1}$ based on estimates for the Galaxy (see \S2), 
$t(z)$ is the age of the universe at the redshift $z$ of the system, 
and $t^*$ is the ``turn-on'' time corresponding to the onset of astration
and Fe production in the system.
It was found that the available data on [Fe/H] for essentially all the DLA 
systems lie between the prompt inventory value and the upper bound obtained
for $t^*=0$. The large dispersion in [Fe/H] at a given $z$ could be 
explained by different $t^*$ for different DLA systems. It was found that 
for the above closed-system model, most systems at a given $z$ have $t^*$ 
in a range close to $t(z)$ (i.e., they ``turned on'' long after the
big bang so that their baryonic matter had to be stored in the IGM for this
long period).

The astrophysical implications of the late turn-on of most DLA systems in 
the above closed-system model were explored by us in a recent study
based on the cold dark matter (CDM) model of hierarchical structure 
formation (Qian \& Wasserburg 2003b). 
We there associated the evolution of a DLA system with the growth 
of a halo through infall of CDM and gas. The late turn-on was reinterpreted 
in terms of two factors. First, 
a halo must achieve some threshold mass before
star formation could occur. As a halo associated with a low-$\sigma$ 
density fluctuation reaches a given mass at a later 
time, Fe production in such a halo would be delayed simply due to the later 
onset of star formation. Second, subsequent to the onset of star formation,
the growth of (Fe/H) in the gas of a halo is determined by the competition 
between production by SNe II and dilution by gas infall. The uniform growth 
with time does not set in until infall ceases. Thus, the late ``turn-on'' 
does not require the introduction of an arbitrary delay time $t^*$ but is a 
direct result of the infall rates in the CDM model of hierarchical structure 
formation.

We here extend our model for chemical evolution of DLA systems to a more 
general discussion of galactic chemical evolution. In \S2 we review the 
consequences of astration concurrent with infall for the chemical evolution 
of gas in a halo that is not disrupted by SNe II. We discuss the effects of 
infall cessation, particularly the resultant metallicity distribution of 
stars in \S3. The connection with the CDM model of hierarchical structure 
formation is described in \S4. A comparison of our chemical evolution model 
with the data on DLA systems is presented in \S5. 
Possible scenarios of globular cluster 
formation in connection with this model are discussed and conclusions given 
in \S6.

\section{Chemical Evolution of Gas in an Accreting Halo without Disruption} 
We first consider the chemical evolution of gas in a
halo that grows according to hierarchical structure formation. We
assume that CDM and gas are fed into the halo at a fixed mass ratio through 
infall. For simplicity, the infalling gas from the IGM is assumed to be pure 
big bang debris (i.e., metal-free). Observations show that the IGM is 
significantly enriched with ``metallicities'' corresponding to 
[Fe/H]~$\sim -3$ over a wide range of redshift $1.5\leq z\leq 5.5$ 
(e.g., Songaila 2001; Pettini et al. 2003; see also 
Simcoe, Sargent, \& Rauch 2002, 2004; Schaye et al. 2003; 
Aguirre et al. 2004). Thus, the assumption of metal-free infalling gas is 
not necessarily valid. However, possible metallicities at the level of 
[Fe/H]~$\sim -3$ for the infalling gas are soon overwhelmed by 
enrichments associated with astration in the halo. Insofar as the evolution
of the halo substantially enhances the metal content of its gas to well 
above [Fe/H]~$\sim -3$, neglecting the metallicity of the infalling
gas is justified.

The history of star formation in the halo is crucial to the
chemical evolution of the gas. The issue of what the first
generation of stars were like has long been a matter of intense
interest. Numerical simulations of collapse and fragmentation of
metal-free gas clouds indicate that the first stars (or baryonic
aggregates) were very massive. Formation of normal stars may not
have occurred until a ``metallicity'' corresponding to
[Fe/H]~$\sim -4$ was reached in the IGM (see Bromm \& Larson 2004,
and references therein). We do not consider the effect of this
threshold metallicity that may possibly be required for normal
astration. We assume that normal astration proceeds
at a constant rate per H atom in the gas once the halo reaches
some critical mass (see \S4). From the discovery of a
low-mass Galactic halo star with [Fe/H]~$=-5.3$ (Christlieb et al.
2002), it is evident that some low-mass stars were formed when the
level of heavy metals in the IGM was extremely low, far below the 
threshold metallicity for normal astration suggested by theoretical
models (Bromm et al. 2001). 

It has long been recognized that stars in the Galactic halo form a
distinct population. Extensive observational work has established that 
these stars have distinctively low [Fe/H] ranging from typically
$\sim -2$ (e.g, Carney et al. 1996) down to $-5.3$ (see Fig. 5 in
Christlieb 2003). There is a clear break at
[Fe/H]~$\sim -1$ observed in the Galactic evolution of [Ba/Fe] and
[O/Fe] with [Fe/H]. For [Ba/Fe], this break is due to the lack of
major $s$-process contributions from asymptotic giant branch
stars to the Ba in the interstellar medium (ISM) for [Fe/H]~$<-1$
(cf., Busso, Gallino, \& Wasserburg 1999), while for [O/Fe], it is
due to the lack of major Fe contributions from SNe Ia for
[Fe/H]~$<-1$ (Tinsley 1979; see also Kobayashi et al. 1998). 
In addition, the abundance patterns of 
the elements from the Fe group down to O observed in Galactic halo 
stars are characteristic of the nucleosynthetic yields of SNe II.
Thus, the overall abundances of ``metals'' in the early Galaxy 
corresponding to [Fe/H]~$<-1$ in the ISM were clearly
dominated by contributions from SNe II with massive ($\sim
10$--$100\,M_\odot$) progenitors. We note that the marked change at
[Fe/H]~$\sim -3$ observed for the abundances of heavy $r$-process
nuclei in Galactic halo stars (e.g., McWilliam et al. 1995; 
Sneden et al. 1996; Westin et al. 2000; Burris et al. 2000) appears 
to require a source other than SNe II for
these nuclei (Qian \& Wasserburg 2003a). However, this $r$-process
source cannot produce any significant amount of the elements from
the Fe group down to O (e.g., Qian \& Wasserburg 2002), and
therefore, does not affect our discussion of the chemical
evolution of these elements here.

Based on the above discussion, we only consider metal production
by SNe II subsequent to the onset of astration during the early
evolution of a halo. The general effects of SNe Ia that contribute mainly 
the Fe group elements at [Fe/H]~$\gtrsim -1$ will be discussed 
in \S3. For convenience, we use Fe as a representative metal element. 
At [Fe/H]~$<-1$, the other elements (from the Fe group down to O)
essentially evolve in proportion to Fe as specified by the effective 
yield pattern of SNe II (see Table 1 in Qian \& Wasserburg 2002).
If gas in the halo is homogeneous and is not expelled by explosions of
SNe II, the basic equation governing its chemical evolution (Qian
\& Wasserburg 2003b) may be written as:
\begin{equation}
\frac{d({\rm Fe/H})}{dt}=\frac{P_{\rm Fe}}{(\rm H)}-\frac{1}{(\rm
H)} \frac{d({\rm H})_{\rm in}}{dt}({\rm Fe/H}).
\label{feh}
\end{equation}
Here (Fe/H) is the number ratio of Fe to H atoms in the gas,
$P_{\rm Fe}$ is the net rate of Fe production by SNe II, (H) is
the number of H atoms in the gas, and $d({\rm H})_{\rm in}/dt$ is
the rate of infall of H atoms into the halo. The time $t=0$
corresponds to the big bang. The equation governing (H) is:
\begin{equation}
\frac{d({\rm H})}{dt}=\frac{d({\rm H})_{\rm as}}{dt}+ \frac{d({\rm
H})_{\rm in}}{dt}, \label{ht}
\end{equation}
where $d({\rm H})_{\rm as}/dt<0$ is the astration rate. During the
infall dominated regime $|d({\rm H})_{\rm as}/dt|\ll d({\rm H})_{\rm in}/dt$
so that ${\rm (H)}\approx {\rm (H)}_{\rm in}$. Then for this regime 
equation (\ref{feh}) can be rewritten as:
\begin{equation}
\frac{dZ_{\rm Fe}}{dt}=\lambda_{\rm Fe}-\lambda_{\rm in}Z_{\rm Fe}, 
\label{zfe}
\end{equation}
where $Z_{\rm Fe}\equiv {\rm (Fe/H)/(Fe/H)}_\odot$,
$\lambda_{\rm Fe}\equiv P_{\rm Fe}/[{\rm (Fe/H)}_\odot({\rm H})]$, and
$\lambda_{\rm in}\equiv d\ln{\rm (H)}_{\rm in}/dt$. Note that for
$Z_{\rm Fe}$ to increase rapidly, $\lambda_{\rm in}Z_{\rm Fe}$ must
be significantly smaller than $\lambda_{\rm Fe}$.

The general solution to equation (\ref{zfe}) for a specific halo is
\begin{equation}
Z_{\rm Fe}(t)=\exp\left[-\int^t_{t_0}\lambda_{\rm in}(t')dt'\right]
\int^t_{t_0}\lambda_{\rm Fe}(t'')\exp\left[\int^{t''}_{t_0}
\lambda_{\rm in}(t''')dt'''\right]dt'',
\label{zfet}
\end{equation}
where $t_0$ is the onset of astration and Fe production [i.e.,
$Z_{\rm Fe}(t)\equiv 0$ for $t\leq t_0$]. The time $t$ can be
related to the mass of the halo at that time based on the density
fluctuation associated with the halo (see Fig. 2a and \S4). Given
the rate of infall $\lambda_{\rm in}$ and the rate of Fe
production per H atom in the gas $\lambda_{\rm Fe}$, $Z_{\rm Fe}(t)$ 
only depends upon the time $t_0$ for onset of astration
and the time $t$. For the infall phase, we explicitly assume that 
$P_{\rm Fe}/{(\rm H)}$ is constant and take 
$\lambda_{\rm Fe} = (30\ {\rm Gyr})^{-1}$. This value of 
$\lambda_{\rm Fe}$ corresponds to the average rate of Fe production
by SNe II in the Galaxy [$\sim 1/3$ of (Fe/H)$_\odot$ over a
period of $\sim 10$ Gyr]. The infall rate $\lambda_{\rm in}$ is not an
arbitrary function of $t$ but is directly
determined from the model of hierarchical structure formation
discussed in Barkana \& Loeb (2001) (see below). The evolution of [Fe/H] 
in halos with infall of big bang debris but without gas expulsion by SNe II 
can then be calculated in a straightforward manner.

The rate of gas infall governs the growth of the baryonic mass of a halo.
We assume that a constant fraction $\Omega_b/\Omega_m$ of the total halo
mass is in baryonic matter, where $\Omega_b$ and $\Omega_m$ are the 
fractions of the present critical density contributed by baryonic and all
matter, respectively. The infall rate is then related to the total halo 
mass $M$ (mostly in CDM) as
\begin{equation}
\lambda_{\rm in}(t)=\frac{d\ln M}{dt}=\frac{d\ln M}{dz}\frac{dz}{dt}.
\label{inf}
\end{equation}
For a flat universe, we have
\begin{equation}
\frac{dz}{dt}=-H_0(1+z)\sqrt{\Omega_m(1+z)^3+\Omega_\Lambda},
\label{dzdt}
\end{equation}
where $H_0=100h$ km s$^{-1}$ Mpc$^{-1}$ is the present Hubble
parameter and $\Omega_\Lambda$ is the fraction of the present critical 
density contributed by the cosmological constant. We take $h=0.7$, 
$\Omega_b=0.045$, $\Omega_m=0.3$, and $\Omega_\Lambda=0.7$.
Based on the model of hierarchical structure formation (see \S4), 
we show $M$ as a function of $z$ for halos associated with $1\sigma$ and 
$2\sigma$ density fluctuations ($1\sigma$ and $2\sigma$ halos),
respectively, in Figure 2a. Using the function $M(z)$ for a specific halo, 
we calculate $\lambda_{\rm in}(t)$ from equations (\ref{inf}) 
and (\ref{dzdt}). Assuming that astration starts (see Fig. 2a and \S4)
when a $1\sigma$ halo reaches $M(z_0)=2.8\times 10^8\,M_\odot$
corresponding to $z_0=3.8$ ($t_0=1.6$ Gyr) or when a $2\sigma$
halo reaches $M(z_0)=8.4\times 10^7\,M_\odot$ corresponding to
$z_0=9.7$ ($t_0=0.49$ Gyr), we numerically integrate equation
(\ref{zfe}) to calculate the evolution of [Fe/H]~$=\log Z_{\rm
Fe}$ as a function of $z$ until $z=1.6$ [$t=4.1$ Gyr,
$M(z)=10^{11}\,M_\odot$] for the $1\sigma$ halo or until $z=4.3$
[$t=1.4$ Gyr, $M(z)=10^{11}\,M_\odot$] for the $2\sigma$ halo. The
results are shown in Figure 2b. If we consider smaller values of
$M(z_0)$, say $M(z_0)\sim 10^5\,M_{\odot}$, then the onset
of astration will be moved to higher redshift. The evolution of [Fe/H]
for a $2\sigma$ halo with $M(z_0)=7.1\times 10^4\,M_\odot$ corresponding 
to $z_0=16.8$ ($t_0=0.23$ Gyr) is shown as the dot-dashed curve in 
Figure 2c, where the case for $M(z_0)=8.4\times 10^7\,M_\odot$ 
corresponding to $z_0=9.7$ is also shown as the solid curve.
It can be seen that the dot-dashed curve merges with the
solid curve at $z\sim 8$. This is because [Fe/H] quickly approaches a 
quasi-steady state subsequent to the onset of astration (see below).

Inspection of equation (\ref{zfe}) suggests that if
$\lambda_{\rm in}(t)\gg |d\ln Z_{\rm Fe}(t)/dt|$, then a quasi-steady
state (QSS) exists such that
\begin{equation}
Z_{\rm Fe}(t)\approx Z_{\rm Fe}^{\rm QSS}(t)\equiv
\lambda_{\rm Fe}/\lambda_{\rm in}(t).
\label{qss}
\end{equation}
The approach of $Z_{\rm Fe}(t)$ to $Z_{\rm Fe}^{\rm QSS}(t)$
may be approximated as
\begin{equation}
Z_{\rm Fe}(t)\approx Z_{\rm Fe}^{\rm QSS}(t) -
Z_{\rm Fe}^{\rm QSS}(t_0)\exp\left[-\int^t_{t_0}
\lambda_{\rm in}(t')dt'\right].
\label{ztoqss}
\end{equation}
For $t>t_0+[\lambda_{\rm in}(t_0)]^{-1}$, the term containing the
exponential in equation (\ref{ztoqss}) is insignificant and
$Z_{\rm Fe}(t)$ approximately equals the QSS value $Z_{\rm
Fe}^{\rm QSS}(t)$ (see Qian \& Wasserburg 2003b). 
When we numerically integrate equation (\ref{zfe}) to calculate the 
evolution of [Fe/H]~$=\log Z_{\rm Fe}$ as a function of $z$,
we always use the actual function $M(z)$ for a specific halo to obtain
the corresponding $\lambda_{\rm in}(t)$. For a simple estimate of 
$\lambda_{\rm in}(t)$, we note that the slopes of the $M(z)$ curves
are $|d\ln M/dz|\sim 2.3$ within a factor of a few for both $1\sigma$ 
and $2\sigma$ halos (see Fig. 2a).
We also note that for $z>1$, $\Omega_\Lambda$ can be ignored in equation
(\ref{dzdt}) to give
\begin{equation}
t\approx \frac{2}{3}\frac{1}{H_0\sqrt{\Omega_m(1+z)^3}} \approx
17(1+z)^{-3/2}\ {\rm Gyr}.
\label{tz}
\end{equation}
Using $|d\ln M/dz|\sim 2.3$ and equations (\ref{inf}) and (\ref{tz}), 
we obtain
\begin{equation}
\lambda_{\rm in}(t)\sim 10({\rm Gyr}/t)^{5/3}\ {\rm Gyr}^{-1}
\sim 0.09(1+z)^{5/2}\ {\rm Gyr}^{-1}
\label{infa}
\end{equation}
as a rough estimate for both $1\sigma$ and $2\sigma$ halos. This gives
for these halos
\begin{equation}
Z_{\rm Fe}^{\rm QSS}(t)\sim (1/300)(t/{\rm Gyr})^{5/3}
\sim 0.4(1+z)^{-5/2},
\label{zfeqa}
\end{equation}
or
\begin{equation}
{\rm [Fe/H]}_{\rm QSS}=\log Z_{\rm Fe}^{\rm QSS}(t)
\sim -2.9-2.5\log\left(\frac{1+z}{10}\right).
\label{fehq}
\end{equation}
The above estimate for [Fe/H]$_{\rm QSS}$ is shown as the solid
curve labeled ``quasi-steady state estimate'' in Figure 2b.

It can be seen from Figure 2b that for both $1\sigma$ and
$2\sigma$ halos, [Fe/H] of the gas initially rises very rapidly
and approaches the QSS value, which ranges from [Fe/H]$_{\rm QSS}\sim -2.8$
at $z=8$ to $-1.6$ at $z=2$. For comparison, the case 
of evolution from an initial metal-free state without infall, 
$Z_{\rm Fe}(t)=\lambda_{\rm Fe}t$ (prior to the onset of 
SNe Ia), is also shown in Figure 2b. This should
give an upper bound to $Z_{\rm Fe}$ for all halos and appears
to be a reasonable bound for the observational data on DLA systems
(Wasserburg \& Qian 2000b; Qian \& Wasserburg 2003b; see Fig. 5 here). 
Compared to the case of 
no infall, the Fe enrichment for $1\sigma$ and $2\sigma$ halos is suppressed
due to dilution by infall by a factor of 
$\sim\lambda_{\rm in}(t)t\sim 10({\rm Gyr}/t)^{2/3}$, 
which ranges from a factor of $\sim 4.5$ for $t=3.3$ Gyr ($z=2$) to a 
factor of $\sim 14$ for $t=0.63$ Gyr ($z=8$).

\section{Effects of Infall Cessation}
We now consider that infall of gas ceases at some time
for a halo and investigate the subsequent chemical evolution of gas 
in the halo. The metallicity distribution of stars formed before and
after infall cessation (IC) is studied. The general effects of SNe Ia on the
chemical evolution of gas and stars in a halo are also discussed at 
the end of this section.

\subsection{Chemical Evolution of Gas}
We first consider the chemical evolution of gas in a halo after 
continuous infall ceases sharply at some time $t_{\rm IC}$.
The governing equations in this case are still equations
(\ref{feh}), (\ref{ht}), and (\ref{zfe}). But now
$d({\rm H})_{\rm in}/dt = 0$, $\lambda_{\rm in}=0$, and the new initial
state has $Z_{\rm Fe}(t_{\rm IC})\approx Z_{\rm Fe}^{\rm QSS}(t_{\rm IC})$
as determined by the prior evolution with infall. Thus, for $t>t_{\rm IC}$
but prior to the onset of SNe Ia (see \S3.3),
\begin{equation}
Z_{\rm Fe}(t)=Z_{\rm Fe}(t_{\rm IC})+\lambda_{\rm Fe}(t-t_{\rm IC}).
\end{equation}
Taking $t_{\rm IC}=4.1$ Gyr ($z_{\rm IC}=1.6$) for a $1\sigma$
halo and $t_{\rm IC}=1.4$ Gyr ($z_{\rm IC}=4.3$) for a $2\sigma$
halo, we show the above evolution of [Fe/H]~$=\log Z_{\rm Fe}$
for these two halos in Figure 2b. It can be seen
that the evolution subsequent to infall cessation departs markedly
from the QSS case. For the $2\sigma$ halo, the evolution clearly
begins to approach the case of no infall. In these calculations, we have 
assumed that infall ceases sharply at time $t_{\rm IC}$ corresponding to 
redshift $z_{\rm IC}$ when the mass of a halo reaches $M(z_{\rm IC})=
10^{11}\,M_\odot$ (see Fig. 2a and \S4). If we consider larger values 
of $M(z_{\rm IC})$, say $M(z_{\rm IC})=10^{12}\,M_{\odot}$, this
will only delay the onset of uniform growth of $Z_{\rm Fe}$ to 
lower redshift. The general evolution of [Fe/H] outlined above will
be unaffected. The evolution of [Fe/H] for a $2\sigma$ halo with
$M(z_{\rm IC})=10^{12}\,M_{\odot}$ corresponding to $t_{\rm IC}=2.3$ Gyr 
($z_{\rm IC}=2.8$) is shown as the dashed curve in Figure 2c.

\subsection{Metallicity Distribution of Stars}
In addition to the effect on the evolution of $Z_{\rm Fe}$
discussed above, the instant cessation of infall has another consequence.
During the infall phase, the amount of gas in a halo rapidly
increases as (H)~$\approx({\rm H})_{\rm in}$ for $|d({\rm H})_{\rm
as}/dt|\ll d({\rm H})_{\rm in}/dt$. Subsequent to instant infall
cessation, (H) steadily decreases due to astration. Thus, (H) is
maximal at time $t_{\rm IC}$. As the rate of star formation
increases with (H), this rate also reaches a maximum at time
$t_{\rm IC}$.

The general history of star formation before and after infall
cessation is reflected by the metallicity distribution of stars in
a halo. This distribution is defined as the number $(N)$ of stars
formed per unit interval of [Fe/H]~$=\log Z_{\rm Fe}$,
\begin{equation}
\frac{dN}{d[{\rm Fe/H}]}=-(\ln 10)\frac{Z_{\rm Fe}}
{({\rm H})_\star}\frac{d({\rm H})_{\rm as}/dt}{dZ_{\rm Fe}/dt},
\label{md}
\end{equation}
where $({\rm H})_\star$ is the average number of H atoms in a star
[note that $d({\rm H})_{\rm as}/dt<0$].

The metallicity distribution of stars formed at $t>t_{\rm IC}$ is
simple to obtain. In this case, $d({\rm H})/dt=d({\rm H})_{\rm
as}/dt$ and $dZ_{\rm Fe}/dt=P_{\rm Fe}/[({\rm Fe/H})_\odot({\rm
H})]$ (cf., eqs. [\ref{ht}] and [\ref{zfe}]). Assuming that the SN II
rate is the same as the rate at which massive SN II progenitors are
formed (the instantaneous recycling approximation), we can relate
$P_{\rm Fe}$ to the astration rate $d({\rm H})_{\rm as}/dt$ as
\begin{equation}
P_{\rm Fe}=-\alpha({\rm Fe/H})_\odot\frac{d({\rm H})_{\rm as}}{dt},
\label{pfeas}
\end{equation}
where $\alpha$ is a positive dimensionless constant. Thus, for
$t>t_{\rm IC}$,
\begin{equation}
\frac{dZ_{\rm Fe}}{dt}=-\frac{\alpha}{\rm(H)}
\frac{d({\rm H})_{\rm as}}{dt}=-\alpha\frac{d\ln({\rm H})}{dt}.
\end{equation}
The above equation can be solved to give
\begin{equation}
{\rm(H)}={\rm(H)}_{\rm IC}\exp
\left[-\frac{Z_{\rm Fe}-Z_{\rm Fe}(t_{\rm IC})}{\alpha}\right],
\label{hic}
\end{equation}
where (H)$_{\rm IC}$ is the number of H atoms in the gas at
$t=t_{\rm IC}$. The metallicity distribution at $t>t_{\rm IC}$ is then
\begin{equation}
\frac{dN}{d[{\rm Fe/H}]}=-(\ln 10)\frac{Z_{\rm Fe}}{({\rm H})_\star}
\frac{d{\rm (H)}}{dZ_{\rm Fe}}=\frac{(\ln 10)}{\alpha}
\frac{{\rm (H)}_{\rm IC}}{({\rm H})_\star}Z_{\rm Fe}\exp
\left[-\frac{Z_{\rm Fe}-Z_{\rm Fe}(t_{\rm IC})}{\alpha}\right].
\label{mdic}
\end{equation}
The above metallicity distribution assumes that only SNe II provide Fe. 
Note that this distribution does not depend on
the specific forms of $d({\rm H})_{\rm as}/dt$ and $P_{\rm Fe}$ as
functions of time but only on the ratio of these two quantities.
This is true of any closed system as recognized by Hartwick (1976)
long ago.

In contrast, the metallicity distribution of stars formed during
the infall phase depends on the specific forms of $d({\rm H})_{\rm
as}/dt$ and $P_{\rm Fe}$. In general, the evolution of $Z_{\rm
Fe}$ and (H) during the infall phase must be solved together as
explicit functions of time from equations (\ref{feh}) and
(\ref{ht}) after $d({\rm H})_{\rm as}/dt$, $P_{\rm Fe}$, and the
infall rate $d({\rm H})_{\rm in}/dt$ are specified for a given
halo. We again assume 
$P_{\rm Fe}/[{\rm (Fe/H)}_\odot{\rm (H)}]=\lambda_{\rm
Fe}= (30\ {\rm Gyr})^{-1}$ as in \S2. Using this assumption and 
equation (\ref{pfeas}) gives
\begin{equation}
\frac{d({\rm H})_{\rm as}}{dt}=-\frac{\lambda_{\rm Fe}}{\alpha}{\rm (H)}.
\label{has}
\end{equation}
As a numerical example of estimating $\alpha$, we assume that SNe II with 
progenitors of $\sim 20$--$25\,M_\odot$ produce $\sim 0.1\,M_\odot$ of Fe 
as in the case of SN 1987A but no other
SNe II produce Fe. For a Salpeter initial mass function 
$dN/dM_\star\propto M_\star^{-2.35}$ over a range of stellar masses
$M_\star\sim 0.1$--$100\,M_\odot$, a total of $\sim 1700\,M_\odot$ of 
stars are formed for each Fe-producing SN II. From equation (\ref{pfeas})
we obtain $\alpha\sim 0.1\,M_\odot/(10^{-3}\times 1700\,M_\odot)\sim 1/17$,
where we have used an Fe mass fraction of $\sim 10^{-3}$ corresponding to
the number abundance (Fe/H)$_\odot$. Noting that the Fe yields of SNe II,
especially the dependence of these yields on the progenitor masses, are 
rather uncertain, we take $\alpha=1/18$ as a nominal value and 
examine the effects of a higher value of $\alpha=1/6$ later. For a halo
with $10^{10}\,M_\odot$ of gas and $\lambda_{\rm Fe}=(30\ {\rm Gyr})^{-1}$,
the astration rate in equation (\ref{has}) is 6 and $2\,M_\odot$ yr$^{-1}$
for $\alpha=1/18$ and 1/6, respectively. These astration rates are typically 
much smaller than the infall rate as assumed in \S2. However, to be more
accurate, we can solve $Z_{\rm Fe}$ and (H) together as functions
of time from equations (\ref{feh}), (\ref{ht}), and (\ref{has}).
These two functions can then be used to obtain the metallicity
distribution at $t<t_{\rm IC}$ from equations (\ref{md}) and
(\ref{has}). The metallicity distribution over the whole time domain
$t<t_{\rm IC}$ and $t>t_{\rm IC}$ is shown for a $2\sigma$ halo
with $\alpha=1/18$, $t_0=0.49$ Gyr ($z_0=9.7$), and instant infall 
cessation at $t_{\rm IC}=1.4$ Gyr ($z_{\rm IC}=4.3$)
in Figure 3a (solid curve). It can be seen that this distribution 
has a sharp peak at [Fe/H]~$=-2.2$ and a broad peak at
[Fe/H]~$=-1.3$. The broad peak occurs in the region $t>t_{\rm IC}$
and is easy to understand from equation (\ref{mdic}). This
equation shows that $dN/d{\rm [Fe/H]}$ has a peak at $Z_{\rm Fe}=\alpha$, 
which corresponds to [Fe/H]~$=\log\alpha=-1.3$ for $\alpha=1/18$.

The sharp peak in the metallicity distribution of stars occurs at 
$t=t_{\rm IC}$, the time of infall cessation. The origin of this peak can be
understood by considering equation (\ref{md}) in the immediate
neighborhood of $t_{\rm IC}$ ($t_{\rm IC}^-<t_{\rm IC}<t_{\rm IC}^+$).
Note that $dZ_{\rm Fe}/dt$ has a discontinuity at $t=t_{\rm IC}$
but $Z_{\rm Fe}$ and $d{\rm (H)}_{\rm as}/dt$ are the same
on both sides of this discontinuity. This gives
\begin{equation}
R_{-/+}\equiv\frac{(dN/d[{\rm Fe/H}])_{t_{\rm IC}^-}}
{(dN/d[{\rm Fe/H}])_{t_{\rm IC}^+}}
=\frac{(dZ_{\rm Fe}/dt)_{t_{\rm IC}^+}}
{(dZ_{\rm Fe}/dt)_{t_{\rm IC}^-}}
=\frac{\lambda_{\rm Fe}}{(dZ_{\rm Fe}/dt)_{t_{\rm IC}^-}}.
\label{ric}
\end{equation}
As the QSS approximation $Z_{\rm Fe}\approx Z_{\rm Fe}^{\rm QSS}$
holds during the infall phase ($t<t_{\rm IC}$), $(dZ_{\rm
Fe}/dt)_{t_{\rm IC}^-}\approx 0$ to zeroth order and $R_{-/+}$ is
singular (see eq. [\ref{zfe}]). To first order,
\begin{equation}
\left(\frac{dZ_{\rm Fe}}{dt}\right)_{t_{\rm IC}^-}=
\left(\frac{dZ_{\rm Fe}^{\rm QSS}}{dt}\right)_{t_{\rm IC}}=
\left[\frac{d(\lambda_{\rm Fe}/\lambda_{\rm in})}{dt}\right]
_{t_{\rm IC}}\sim\frac{\lambda_{\rm Fe}}{6}
\left(\frac{t_{\rm IC}}{\rm Gyr}\right)^{2/3},
\label{dzfeic}
\end{equation}
where we have used the estimate for $\lambda_{\rm in}$ in equation
(\ref{infa}) for $1\sigma$ and $2\sigma$ halos to obtain the last result. 
It follows from equations
(\ref{ric}) and (\ref{dzfeic}) that the ratio $R_{-/+}\sim 6({\rm
Gyr}/t_{\rm IC})^{2/3}$ for $1\sigma$ and $2\sigma$ halos is much greater
than unity for a wide range of $t_{\rm IC}$. This leads to a
sudden drop in $dN/d{\rm [Fe/H]}$ at $t=t_{\rm IC}$ and produces a
sharp peak in the metallicity distribution. The value of [Fe/H] at
the sharp peak depends on $t_{\rm IC}$. For a $2\sigma$ halo with
$t_{\rm IC}=1.4$ Gyr ($z_{\rm IC}=4.3$), this peak is at 
[Fe/H]~$\sim\log Z_{\rm Fe}^{\rm QSS}(t_{\rm IC}) \sim -2.2$ 
(see eq. [\ref{zfeqa}] and Fig. 3a).

To compare the effect of instant infall cessation as assumed above
with that of a smooth transition for a $2\sigma$ halo, 
we have calculated models where
$d{\rm (H)}_{\rm in}/dt$ decays smoothly starting at time $t_{\rm
IC}$ with the functional form $\exp[-(t-t_{\rm IC})/\tau]$ for
$t>t_{\rm IC}$. The results are shown in Figure 3a for $\tau =
0.1$ and 0.2 Gyr. As can be seen, the main effect of changing
$\tau$ from 0 (instant infall cessation) to a few tenths of a Gyr
is to increase the fraction of stars formed at $t>t_{\rm IC}$.
This is because there is more gas available for star formation in
the case where the infall rate decays smoothly over an extended
time. However, the width of the sharp peak is increased only
slightly from $\sim 0.06$ dex for $\tau = 0$ to $\sim 0.14$ dex
for $\tau = 0.2$ Gyr. Concerning the total population of stars in the
metallicity distribution, we note that only 10\% of all the stars are in
the sharp peak for $\tau=0$ and this fraction is even smaller for
a finite $\tau$. For a slowly-decaying infall rate with $\tau\sim
1$ Gyr, the sharp peak at $t=t_{\rm IC}$ is still present but
contains only $\sim 2\%$ of the stars and becomes a small feature
of an otherwise smooth metallicity distribution (see Fig. 3b for
the case of $\tau=1$ Gyr).

The bimodal form of the metallicity distribution of stars
discussed above applies to all halos in general. One critical
aspect of this bimodal behavior is the value of [Fe/H] at the
sharp peak. This value is governed by the time $t_{\rm IC}$ for
infall cessation, which depends on the density fluctuation
associated with a halo (see \S4). The metallicity distributions for
a $1\sigma$ halo with instant infall cessation ($\tau=0$) and smooth 
transition to exponential decay of infall ($\tau=1$ Gyr) 
at a relatively late time of $t_{\rm IC}=4.1$ Gyr ($z_{\rm IC}=1.6$) 
are shown in Figure 3c [$\alpha=1/18$ and $t_0=1.6$ Gyr ($z_0=3.8$)
are used]. Compared 
with Figures 3a and 3b for a $2\sigma$ halo with $t_{\rm IC}=1.4$ Gyr 
($z_{\rm IC}=4.3$), the sharp peak for the $1\sigma$ halo is shifted 
to a higher [Fe/H] of $-1.8$. To obtain a peak with a lower [Fe/H] of 
$-3$ would require
$t_{\rm IC}\sim 0.5$ Gyr corresponding to $z_{\rm IC}\sim 10$ (see
eq. [\ref{fehq}]). Comparison of Figures 3a and 3b with 3c also 
shows that the sharp peak for the $2\sigma$ halo contains a much smaller
fraction of stars than that for the $1\sigma$ halo. 
Consider the case of instant infall cessation as a specific example.
For the $2\sigma$ halo, the infall rate at high redshift corresponding to
$t<t_{\rm IC}$ is large compared to the astration rate and only
10\% of stars are formed during the infall phase. In contrast, for
the $1\sigma$ halo, the astration
rate is no longer small compared to the infall rate at lower
redshift near the end of the infall phase and $28\%$ of stars are
formed during this phase. From the model presented here it follows
that galaxies in general should have a bimodal distribution of
stars with a relatively small number of stars defining a sharp
peak at low [Fe/H]. This peak reflects the state when gas infall
ceased or greatly decreased. Note that we have not addressed the
location of stars in a galaxy when discussing their metallicity
distribution. The bimodal metallicity distribution is more likely
to describe stars in the halo than those in the disk of a galaxy.

As discussed above, the metallicity distribution of stars in a halo
is calculated by specifying $\lambda_{\rm Fe}$, $\alpha$, $t_0$, 
$t_{\rm IC}$, and the form of infall cessation. The infall rate 
$\lambda_{\rm in}$ used in the calculation for $t<t_{\rm IC}$ is 
determined from the model of hierarchical struture formation as in \S2. 
The effects of the form of infall cessation (instant or exponential decay)
on the metallicity distribution have been discussed above. To examine
the effects of $\alpha$ and $t_{\rm IC}$, we have calculated the
metallicity distribution for a $2\sigma$ halo with $\alpha=1/6$,
$t_0=0.49$ Gyr ($z_0=9.7$), and instant infall cessation at
$t_{\rm IC}=2.3$ Gyr ($z_{\rm IC}=2.8$). This is shown as the solid
curve in Figure 3d. Compared to the solid curve in Figure 3a 
[$\alpha=1/18$ and $t_{\rm IC}=1.4$ Gyr ($z_{\rm IC}=4.3$)], the sharp
peak in Figure 3d is shifted to 
[Fe/H]~$=-1.9\sim\log Z_{\rm Fe}^{\rm QSS}(t_{\rm IC})$
due to a later $t_{\rm IC}$ and the broad peak (solid curve) in Figure 3d
is shifted to [Fe/H]~$=\log\alpha=-0.8$ due to 
a larger $\alpha$. In addition, $15\%$ of all the stars 
represented by the solid curve in Figure 3d have 
$-0.5\lesssim {\rm [Fe/H]}\lesssim 0$ while the corresponding fraction 
is only $0.3\%$ for the solid curve in Figure 3a. For completeness, 
we note that $t_0$ for the onset of astration has little effect on 
the metallicity distribution as most of the stars are 
formed close to $t=t_{\rm IC}$ for $t<t_{\rm IC}$.

It is evident from inspection of Figure 3 that if a
protogalactic halo were to expel its gas immediately following
the instant cessation of infall, an assemblage of stars with nearly 
identical [Fe/H] would be left behind. This assemblage of stars may be 
accreted by a larger system and become a globular cluster of the larger 
system. In this model, the 
sharp distribution of [Fe/H] for the stars in a globular cluster would
not be the result of a rapid burst of star formation but rather 
the result of the following two factors. First, for a protogalactic halo 
with ongoing infall and astration over an extended period, the rate of 
star formation increases rapidly with time and reaches a maximum 
at the time $t_{\rm IC}$ for instant infall cessation when there is the 
largest amount of gas in the halo. This causes nearly all the 
stars formed during the infall phase to have the metallicity of the gas 
at $t=t_{\rm IC}$. Second, the rapid loss of gas subsequent to infall 
cessation terminates further astration. Our model relates the value of 
[Fe/H] for the globular cluster to $t=t_{\rm IC}$, and therefore, can be 
used to estimate the formation time (relative to the big bang) of the 
cluster based on its [Fe/H] if the initial metallicity of the infalling 
gas is negligible. 

\subsection{General Effects of SNe Ia}
To illustrate the general effects of SNe Ia, we assume that these supernovae
start to contribute Fe when [Fe/H]~$=-1$ is reached at time $t^\dagger$. 
The rate of Fe production by SNe Ia is taken to be twice that for SNe II 
so that a solar Fe abundance can be achieved over a period of $\sim 10$ 
Gyr as in the Galaxy. The total Fe production rate per H atom in 
the gas then increases from $\lambda_{\rm Fe}$ at $t<t^\dagger$ to 
$3\lambda_{\rm Fe}$ at $t>t^\dagger$. Consider halos with instant infall
cessation occurring at $t=t_{\rm IC}$ prior to the onset of SNe Ia (i.e.,
$t^\dagger>t_{\rm IC}$). The evolution of $Z_{\rm Fe}$ at $t>t^\dagger$ 
for these halos is given by
\begin{equation}
Z_{\rm Fe}(t)=Z_{\rm Fe}(t^\dagger)+3\lambda_{\rm Fe}(t-t^\dagger),
\label{zia}
\end{equation}
where $Z_{\rm Fe}(t^\dagger)=0.1$ ([Fe/H]~$=-1$) is taken to correspond 
to the onset of SNe Ia. The above evolution of [Fe/H]~$=\log Z_{\rm Fe}$ 
is shown for a $1\sigma$ halo with $t_0=1.6$ Gyr ($z_0=3.8$) and 
$t_{\rm IC}=4.1$ Gyr ($z_{\rm IC}=1.6$) and for a $2\sigma$ halo with
$t_0=0.49$ Gyr ($z_0=9.7$) and $t_{\rm IC}=1.4$ Gyr ($z_{\rm IC}=4.3$)
in Figure 2b. The evolution of [Fe/H] at $t>t^\dagger$ for the case of 
no infall is also given by equation (\ref{zia}) and shown in Figure 2b. 
The only effect of SNe Ia on the evolution 
of [Fe/H] for all halos is to increase the rate of Fe enrichment at 
$t>t^\dagger$.

As in the case of infall cessation, the sudden increase of the Fe 
production rate per H atom in the gas at the onset of SNe Ia leads to
a sharp drop of $dN/d{\rm [Fe/H]}$ at [Fe/H]~$=-1$. This is shown for
a $2\sigma$ halo with $\alpha=1/18$ and smooth transition to exponential
decay of infall ($\tau=1$ Gyr) at $t_{\rm IC}=1.4$ Gyr ($z_{\rm IC}=4.3$)
in Figure 3b and for another $2\sigma$ halo with $\alpha=1/6$ and instant
infall cessation at $t_{\rm IC}=2.3$ Gyr ($z_{\rm IC}=2.8$) in Figure 3d.
For the cases where only SNe II provide Fe,
the metallicity distribution of stars has a broad peak at 
[Fe/H]~$=\log\alpha=-1.3$ for Figure 3b and $-0.8$ for Figure 3d 
(solid curves). If we include SN Ia contributions for
[Fe/H]~$>-1$, this broad peak is truncated and a new broad peak is formed 
at [Fe/H]~$=\log(3\alpha)=-0.8$ for Figure 3b and $-0.3$ for Figure 3d
(dashed curves). This change to the metallicity distribution
is caused by the more rapid increase of [Fe/H] with additional 
contributions from SNe Ia. The above effect of SNe Ia on the metallicity 
distribution of stars should apply to all halos. For completeness, we
note that the final halo mass for the $2\sigma$ halo shown in Figure 3b
is $5.5\times 10^{11}\,M_\odot$ corresponding to a baryonic mass of
$8.3\times 10^{10}\,M_\odot$. In this case $89\%$ of the baryonic matter
has been consumed in star formation prior to the onset of SNe Ia. In
comparison, the final halo mass for the $2\sigma$ halo shown in Figure 3d 
is $10^{12}\,M_\odot$ corresponding to a baryonic mass of
$1.5\times 10^{11}\,M_\odot$. In this case $45\%$ of the baryonic matter
has been consumed in star formation prior to the onset of SNe Ia.

\section{Cosmological Input for Halo Evolution}
In the formal discussion of \S2 and \S3 on the chemical evolution of a halo,
we used the times for the onset of astration ($t_0$) and infall cessation
($t_{\rm IC}$) and the infall rate ($\lambda_{\rm in}$), all of which are
related to the growth of the halo as determined by the CDM model of 
hierarchical structure formation. We now explicitly explore how this
cosmological input is determined. The physical conditions governing the 
general evolution of a halo are closely related to the mass (mostly in CDM)
of the halo. For a halo associated with an $n\sigma$ density fluctuation 
(an $n\sigma$ halo), its mass $M$ at redshift $z>1$ can be estimated from
\begin{equation}
n\sigma(M)\approx 1.33(1+z),
\label{nsig}
\end{equation}
where $\sigma(M)$ is the standard deviation of the present power spectrum
of overdensity on mass scale $M$. Using the version of equation (\ref{nsig})
presented in the Appendix and taking $\sigma(M)$ from
Figure 5 of Barkana \& Loeb (2001), we show $M$ as a function of $z$ for 
$1\sigma$ and $2\sigma$ halos, respectively, in Figure 2a. The function 
$M(z)$ for a specific halo was used to calculate the infall rate 
$\lambda_{\rm in}$ from equations (\ref{inf}) and (\ref{dzdt}) 
in \S2. In addition, when discussing the effects of infall cessation in \S3, 
we assumed that infall ceases at time $t_{\rm IC}$ corresponding to redshift 
$z_{\rm IC}$ at which the mass of a halo reaches $M(z_{\rm IC})=M_{\rm IC}$. 
For $M_{\rm IC}=10^{11}\,M_\odot$, $z_{\rm IC}=1.6$ or 4.3 for a $1\sigma$ 
or $2\sigma$ halo, respectively (see Figure 2a). If 
$M_{\rm IC}=10^{12}\,M_{\odot}$ is assumed, then infall cessation is delayed
to $z_{\rm IC}=0.8$ or 2.8 for a $1\sigma$ or $2\sigma$ halo, respectively.
The redshifts at which infall cessation occurs for other $n\sigma$ halos are 
given for $M_{\rm IC}=10^{11}\,M_{\odot}$ and $10^{12}\,M_{\odot}$, 
respectively, by the corresponding curves in Figure 4.
The range of $M_{\rm IC}=10^{11}$ to $10^{12}\,M_{\odot}$ is in accord with
the halo masses of typical galaxies.

A critical physical parameter for the onset of astration in a halo is the 
virial temperature $T_{\rm vir}$ of its gas. For $z>1$ this temperature
is related to the halo mass $M$ as
\begin{equation}
T_{\rm vir}\approx 1.04\times 10^4\left(\frac{\mu}{0.6}\right)
\left(\frac{M}{10^8\,M_\odot}\right)^{2/3}
\left(\frac{1+z}{10}\right)\ {\rm K},
\label{tvir}
\end{equation}
where $\mu$ is the mean molecular weight and $\mu=1.22$ or 0.6 for
a neutral or ionized gas of primordial composition, respectively.
Following Barkana \& Loeb (2001), we assume that astration starts 
in a halo when the gas reaches a threshold virial temperature 
$T_{\rm vir,0}$. At a given redshift $z$, the mass $M=M_0$ that is 
required for a halo to reach $T_{\rm vir}=T_{\rm vir,0}$
can be obtained from equation (\ref{tvir}). Using this equation 
(see Appendix), we show $M_0$ as a function of $z$ for $T_{\rm vir,0}=300$ K
($\mu=1.22$) and $T_{\rm vir,0}=10^4$ K ($\mu=0.6$), respectively, 
in Figure 2a. The redshift $z=z_0$ at which the mass $M(z)$ of an 
$n\sigma$ halo matches $M_0(z)$ 
then corresponds to the onset of astration ($t_0$) in this halo. For
$T_{\rm vir,0}=10^4$ K, $z_0=3.8$ or 9.7 for a $1\sigma$ or $2\sigma$ halo, 
respectively (see Figure 2a). If $T_{\rm vir,0}=300$ K
is assumed, then the onset of astration occurs earlier at
$z_0=7.3$ or 16.8 for a $1\sigma$ or $2\sigma$ halo, respectively.
The redshifts at which astration starts for other $n\sigma$ halos
are given for $T_{\rm vir,0}=300$ or $10^4$ K, respectively, by the 
corresponding curves in Figure 4.

The value of $T_{\rm vir,0}$ for the onset of astration depends crucially
on the available coolant. In the presence of cooling by H$_2$ molecules, 
$T_{\rm vir,0}=300$ K is appropriate. If H$_2$ molecules are dissociated 
(e.g., Couchman \& Rees 1986; Haiman, Rees, \& Loeb 1997; 
Ciardi, Ferrara, \& Abel 2000), then $T_{\rm vir,0}=10^4$ K is required
for efficient cooling by atomic species. While the choice of 
$T_{\rm vir,0}=300$ or $10^4$ K does not affect the long-term chemical
evolution of a halo (see Fig. 2c), it can make an important difference
in the early evolution of the halo. This difference concerns the possibility
for a single SN II to expel all the gas, thereby disrupting the chemical 
evolution of a halo. This possibility can be assessed by comparing
the explosion energy of $\sim 10^{51}$ erg for an SN II with the 
binding energy $E_{\rm bi,gas}$ of the gas in the halo. For $z>1$ this 
binding energy is related to the halo mass $M$ as
\begin{equation}
E_{\rm bi,gas}\approx 4.31\times 10^{52}
\left(\frac{M}{10^8\,M_\odot}\right)^{5/3}
\left(\frac{1+z}{10}\right)\ {\rm erg}.
\label{ebgas}
\end{equation}
We consider a simple scenario where all the gas is expelled 
when $E_{\rm bi,gas}$ for a halo is less than $10^{51}$ erg.
Using equation (\ref{ebgas}) (see Appendix), we show the halo mass 
$M=M_{\rm bi}$ for $E_{\rm bi,gas}=10^{51}$ erg as a function of $z$ 
in Figure 2a. The redshifts at which different $n\sigma$ halos reach
$E_{\rm bi,gas}=10^{51}$ erg are given by the corresponding curve in 
Figure 4. As can be seen from this figure, if $T_{\rm vir,0}=300$ K
is assumed for the onset of astration, then the chemical evolution of 
all halos will be disrupted by SNe II until $E_{\rm bi,gas}=10^{51}$ erg 
is reached. In contrast, for $T_{\rm vir,0}=10^4$ K, such disruption cannot 
occur as $E_{\rm bi,gas}=10^{51}$ erg is always reached prior to the onset
of astration.

\section{Comparison with Data on DLA Systems}
We now compare our chemical evolution model with the data on [Fe/H] for
DLA systems. Using $T_{\rm vir,0}=10^4$ K for the onset of astration, 
$M_{\rm IC}=10^{11}\,M_\odot$ for instant infall cessation, and
[Fe/H]~$=-1$ for the onset of SNe Ia, we show the evolution of [Fe/H]
as a function of $z$ for $0.67\sigma$, $1\sigma$, and $2\sigma$ halos
along with the data on DLA systems (Prochaska et al. 2003) in Figure 5
(see also Qian \& Wasserburg 2003b).
The case of no infall is also shown in this figure. It can be seen that
the chemical evolution model presented here provides a good description
of the data. For any given $z$ the model gives a wide range of [Fe/H] 
corresponding to three different evolutionary stages of DLA systems:
sharp initial rise after the onset of astration, quasi-steady state due 
to competition between Fe production and dilution by infall, and uniform
growth subsequent to infall cessation. The evolutionary stage of a DLA
system at a given $z$ depends on the density fluctuation associated with
the system. The large dispersion in [Fe/H] for DLA systems at any given 
$z$ is then accounted for by the different density fluctuations associated 
with these systems. A detailed quantitative assessment of the model would
require many more DLA systems with accurate measurements of [Fe/H] at a 
given $z$ (see Qian \& Wasserburg 2003b for an illustrative example). 
The model also provides an explanation 
for the baseline enrichment of [Fe/H]~$\sim -3$ observed for DLA systems.
Subsequent to the onset of astration, it takes only 
$\sim 30$ Myr to reach [Fe/H]~$\sim -3$ for 
$\lambda_{\rm Fe}=(30\ {\rm Gyr})^{-1}$. Due to this rapid initial rise 
of [Fe/H], the probability to find DLA systems with [Fe/H] below $\sim -3$
is very low. The baseline enrichment of [Fe/H]~$\sim -3$ for DLA systems
is similar to the abundance level in the IGM that was attributed to a
prompt inventory of metals suggested by other arguments 
(e.g., Wasserburg \& Qian 2000a; Qian \& Wasserburg 2002; Qian, Sargent,
\& Wasserburg 2002). Based on the model presented here, this postulated 
prompt inventory is not required to explain the data on DLA systems 
although it is also not in conflict with these data.

\section{Discussion and Conclusions}
We presented a model for the chemical evolution of the gas
and stars in halos as a function of redshift. This model assumes
that a normal population of stars including SN II progenitors can
form from big bang debris and that SNe II will provide metals such
as the elements from the Fe group down to O. If there is a prompt inventory
of metals corresponding to [Fe/H]~$\lesssim -3$ in the IGM, this will simply
provide a baseline enrichment for the initial state and will not alter the
results. The infall of gas into 
a halo is assumed to closely follow that of CDM. The infall rate and
the general evolution of the halo are prescribed by the model of 
hierarchical structure formation based on the density fluctuation 
associated with the halo. For systems that are not disrupted by SN II 
explosions, the chemical evolution of the gas and stars is explicitly 
determined by the infall rate as compared with the astration rate and 
the corresponding rate of metal production by SNe II per H atom in the gas. 
Once the astration rate and the metal production rate are specified,
the only ``free'' parameters in this model are the times (or states) at 
which astration starts and infall ceases for a halo.
For all systems not disrupted by SN II explosions, if there is a 
cessation of gas infall, the metallicities of stars follow a bimodal 
distribution. This distribution is characterized by a sharp peak at the 
value of [Fe/H] corresponding to the time of infall cessation and by a
broad peak at a higher value of [Fe/H] corresponding to the
subsequent period of astration during which the bulk of the
remaining gas forms stars. The fraction of stars in the sharp peak
depends on how rapidly the infall cessation takes place. While it
is not clear how observational bias may affect the proper
metallicity distribution of stars, we note that the distribution
observed for the Galactic halo stars (e.g., Fig. 5 in Christlieb 2003) 
and the bimodal distribution presented here show a strong resemblance.
Based on our model, a distinct narrow peak at low [Fe/H] for the
Galactic halo stars would signify the time when infall started to
decrease for the Galaxy. We also discussed the general effects of SNe Ia 
on the chemical evolution of the gas and stars in halos. For 
[Fe/H]~$\gtrsim -1$ the production rate for the Fe group elements 
is increased due to additional contributions from SNe Ia. Consequently,
the broad peak in the metallicity distribution of stars corresponding to
the case where only SNe II provide Fe is truncated at [Fe/H]~$\sim -1$ 
and a new broad peak is formed at a higher [Fe/H] value.

We also considered the physical conditions governing the
evolution of a halo. The mass $M$ of a halo is an important
parameter and its growth as a function of redshift is determined
by the model of hierarchical structure formation based on the
$n\sigma$ density fluctuation associated with the halo. We
assumed that the cessation of infall for both baryonic and
dark matter corresponds to the redshift at which the halo mass
reaches $M_{\rm IC}=10^{11}$ to $10^{12}\,M_\odot$. This range of 
$M_{\rm IC}$ is in accord with halo masses of typical galaxies. The other 
two important parameters of a halo are the virial temperature $T_{\rm vir}$ 
and the binding energy $E_{\rm bi,gas}$ of the gas. We assumed that
astration starts in a halo when a critical virial temperature
$T_{\rm vir,0}$ corresponding to a halo mass $M_0$ is reached. In the 
presence of cooling by H$_2$ molecules, $T_{\rm vir,0}=300$ K is appropriate 
and $M_0$ ranges from $\sim 10^5$ to $10^6\,M_\odot$ for different 
$n\sigma$ halos. If H$_2$ molecules are dissociated, 
then $T_{\rm vir,0}=10^4$ K is required and $M_0$ ranges from 
$\sim 10^8$ to $10^9\,M_\odot$. While the assumed
value of $T_{\rm vir,0}$ does not affect the long-term chemical
evolution of a halo, it makes an important difference concerning the
effect of SN II explosions on the early evolution of the halo. 
For $T_{\rm vir,0}=10^4$ K, astration starts in a halo after the halo mass
reaches $M_{\rm bi}$ corresponding to a gas binding energy of 
$E_{\rm bi,gas}=10^{51}$ erg. In this case, the chemical evolution of the
gas and stars in the halo will not be disrupted by SN II explosions and
is summarized in the preceding paragraph. In contrast, for 
$T_{\rm vir,0}=300$ K, astration starts in a halo before 
the halo mass reaches $M_{\rm bi}$ and all low-mass halos with
$M_0<M<M_{\rm bi}$ will have their gas dispersed into the IGM by a single 
SN II. If such low-mass halos indeed form, then
their gas expulsion by SNe II may significantly
enrich the IGM at very high redshift. For example, at $z=10$, halos with
$M_0<M<M_{\rm bi}$ are associated with 1.3--1.7$\sigma$ density fluctuations
(see Fig. 4) and a fraction 
$F(M_0<M<M_{\rm bi})=\sqrt{(2/\pi)}\int_{1.3}^{1.7}\exp(-x^2/2)dx
\approx 10\%$ of all baryonic
matter is being processed through such halos. Taking 
$Z_{\rm Fe}\sim Z_{\rm Fe}^{\rm QSS}\sim 10^{-3}$ (see eq. [\ref{zfeqa}])
for the gas expelled from such halos, we estimate that their gas expulsion 
would result in an average metallicity of $Z_{\rm Fe}^{\rm IGM}\sim 
F(M_0<M<M_{\rm bi})Z_{\rm Fe}^{\rm QSS}\sim 10^{-4}$
corresponding to [Fe/H]$_{\rm IGM}\sim -4$ in the IGM at $z=10$.
We expect that $T_{\rm vir,0}$ for the onset of astration would increase
from 300 to $10^4$ K as various radiation sources turned on during the
evolution of the universe. So the range of low-mass halos that would be
disrupted by SN II explosions would decrease with time and eventually
vanish. The question of whether such low-mass halos may actually provide
significant metal enrichment of the IGM at very high redshift will be 
discussed in a subsequent paper that considers aspects 
regarding dissociation of H$_2$ molecules.

We proposed a possible mechanism of globular cluster formation based on
the sharp peak in the metallicity distribution of stars. This peak 
corresponds to the time of infall cessation. If infall ceases rather
suddenly for a protogalactic halo and its gas is rapidly expelled
afterwards, an assemblage of stars with nearly identical [Fe/H] will be
left behind. This assemblage of stars may be accreted by a larger system
and become a globular cluster of the larger system. 
To estimate the masses and metallicities of such globular 
clusters, we assume $T_{\rm vir,0}=10^4$ K for the onset of astration.
In this case the gas in a halo will be enriched by SNe II without disruption
subsequent to the onset of astration. During the infall phase the amount of 
infalling gas increases rapidly on a timescale of 
$\lambda_{\rm in}^{-1}\sim 0.04[10/(1+z)]^{5/2}$ Gyr (see eq. [\ref{infa}]).
For an astration rate of
$d({\rm H})_{\rm as}/dt=-(\lambda_{\rm Fe}/\alpha)({\rm H})$ with
$\lambda_{\rm Fe}=(30\ {\rm Gyr})^{-1}$ and $\alpha=1/18$,
the fraction of the halo mass $M$ in stars at redshift $z$ is $f_\star$
and can be estimated
as $f_\star\sim (\Omega_b/\Omega_m)(\lambda_{\rm Fe}/\alpha)
\lambda_{\rm in}^{-1}\sim 4\times 10^{-3}[10/(1+z)]^{5/2}$. 
If infall ceases at redshift $z$,
then the sharp peak in the metallicity distribution contains a total mass
$\sim f_\star M$ of stars with essentially the same [Fe/H] as given 
approximately by [Fe/H]$_{\rm QSS}$ for this $z$ (see eq. [\ref{fehq}]).
The assumed value of $T_{\rm vir,0}=10^4$ K corresponds to
$M_0=8.1\times 10^7$ and $2.6\times 10^8\,M_\odot$ for $z=10$ and 4, 
respectively (see Fig. 2a). So in the above scenario, the globular clusters 
would have masses exceeding $f_\star M_0\sim 3\times 10^5\,M_\odot$ and 
[Fe/H]~$\sim -3$ if they were formed at $z=10$ while they would have masses 
exceeding $f_\star M_0\sim 6\times 10^6\,M_\odot$ and 
[Fe/H]~$\sim -2.1$ if they were formed at $z=4$. No small globular clusters 
with masses of $\sim 10^4\,M_\odot$ are formed in this
scenario unless the larger aggregates are fragmented. We note that the 
masses derived for Galactic globular clusters are rather uncertain and
range from $\sim 300$ to $\sim 10^7\,M_\odot$ with a mean of
$\sim 2\times 10^5\,M_\odot$ (Mandushev, Spassova, \& Staneva 1991;
Richer et al. 1991). The metallicity distribution
of Galactic globular clusters is bimodal with peaks at [Fe/H]~$\sim -1.6$ 
and $-0.5$, respectively (Armandroff \& Zinn 1988). Approximately 10\% of
the clusters have [Fe/H]~$<-2$. The masses and metallicities of globular 
clusters in our model are broadly consistent with those of the Galactic
population. A detailed comparison requires considerations of possible
ranges for $\lambda_{\rm Fe}$, $\lambda_{\rm in}$, and $\alpha$.
We also note that stars in an individual Galactic globular cluster
typically have a very narrow range of [Fe/H], which can be explained by
our model. However, there is a considerable range 
in [Fe/H] for the most massive Galactic globular cluster $\omega$
Centauri (Suntzeff \& Kraft 1996; Norris, Freeman, \& Mighell
1996) with a mass of $7\times 10^6\,M_\odot$ (Richer et al. 1991). The
range of [Fe/H] for $\omega$ Centauri may have resulted from further 
astration subsequent to infall cessation based on our model. As an SN II 
cannot unbind the gas in the protogalactic
halo associated with such a large cluster 
and some additional mechanism for gas loss must be involved, a more
extensive evolution of [Fe/H] would be expected. The gas loss may have to 
occur by e.g., passing through the Galactic disk. 

If we assume $T_{\rm vir,0}=300$ K for the onset of astration, then a 
different scenario of globular cluster formation may occur.
This scenario is possible if H$_2$ molecules are not dissociated and 
the IGM is not reionized so we here only consider it for very high redshift, 
say $z\sim 10$. For $z=10$, $M_0=1.5\times 10^5\,M_\odot$ for the assumed 
value of $T_{\rm vir,0}=300$ K and $M_{\rm bi}=9.9\times 10^6\,M_\odot$
(see Fig. 2a). A low-mass halo with $M_0<M<M_{\rm bi}$
has started astration but its gas will be expelled by a single SN II.
As the metals produced by the SN II are expelled along with the gas, 
any metals in the stars already formed cannot be due to this SN II
but must have come from the infalling gas. In other words, the metallicities
of these stars represent the metal inventory of [Fe/H]~$\lesssim -3$
provided to the IGM at $z\sim 10$ by sources at even higher redshift.
The assemblage of stars after the SN II explosion resembles 
a globular cluster with a mass of $\sim f_\star M_0\sim 500\,M_\odot$ to 
$\sim f_\star M_{\rm bi}\sim 3\times 10^4\,M_\odot$. We note that
Galactic globular clusters in this mass range typically have [Fe/H]
much higher than $-3$. Whether globular clusters in the same mass range 
but with [Fe/H]~$\lesssim -3$ exist remains to be seen. The possibility
that low-mass halos might form at $z<10$ from an enriched IGM also
remains to be explored.

\acknowledgments
We would like to acknowledge the anonymous referee for one of our
earlier studies on cosmochemical evolution (Qian \& Wasserburg 2003a)
for being supportive and for suggesting that we might pay attention to
infall. Good advice sometimes does fall in. We also want to thank the
anonymous referee for the present work and the Scientific Editor,
Brad Gibson, for helpful suggestions. This work was supported
in part by DOE grants DE-FG02-87ER40328, DE-FG02-00ER41149 (Y. Z. Q.)
and DE-FG03-88ER13851 (G. J. W.), Caltech Division Contribution
8909(1111).

\appendix
\section{Equations for Halo Evolution}
In this appendix, we summarize the equations that are used to
calculate the mass $M$, the virial temperature $T_{\rm vir}$,
and the gas binding energy $E_{\rm bi,gas}$ of a halo for any
redshift $z$. While these equations were discussed in many
works, our discussion here closely follows that of 
Barkana \& Loeb (2001). We consider a flat universe with 
$\Omega_m=0.3$, $\Omega_b=0.045$, $\Omega_\Lambda=0.7$, and $h=0.7$. 

The evolution of the mass $M$ for an $n\sigma$ halo is prescribed by
\begin{equation}
1.686\frac{D(0)}{D(z)}=n\sigma(M),
\label{d0z}
\end{equation}
where $D(z)$ is the growth factor (Peebles 1980) and $\sigma(M)$ is
the standard deviation of the present power spectrum of overdensity
on mass scale $M$. The growth factor can be approximated as
(e.g., Eisenstein \& Hu 1998)
\begin{equation}
D(z)\propto\frac{\Omega_m^z(1+z)^{-1}}{[\Omega_m^z]^{4/7}
-\Omega_\Lambda^z+[1+(\Omega_m^z/2)][1+(\Omega_\Lambda^z/70)]},
\label{gf}
\end{equation}
where
\begin{eqnarray}
\Omega_m^z&=&\frac{\Omega_m(1+z)^3}{\Omega_m(1+z)^3+\Omega_\Lambda},\\
\Omega_\Lambda^z&=&\frac{\Omega_\Lambda}{\Omega_m(1+z)^3+\Omega_\Lambda}.
\end{eqnarray}
Note that the proportionality relation in equation (\ref{gf}) is 
sufficient to determine the ratio $D(0)/D(z)$ in 
equation (\ref{d0z}). For $z>1$, equation (\ref{d0z}) can be simplified
to give equation (\ref{nsig}) for the adopted cosmological parameters.

The virial temperature $T_{\rm vir}$ of the gas in a halo of mass $M$
at redshift $z$ is
\begin{equation}
T_{\rm vir}=1.98\times 10^4\left(\frac{\mu}{0.6}\right)
\left(\frac{M}{10^8h^{-1}\,M_\odot}\right)^{2/3}
\left(\frac{\Omega_m}{\Omega_m^z}\frac{\Delta_c}{18\pi^2}\right)^{1/3}
\left(\frac{1+z}{10}\right)\ {\rm K},
\label{tvirz}
\end{equation}
where $\Delta_c=18\pi^2-82\Omega_\Lambda^z-39(\Omega_\Lambda^z)^2$
(cf. Bryan \& Norman 1998).
The binding energy $E_{\rm bi,gas}$ of the gas is
\begin{equation}
E_{\rm bi,gas}=5.45\times 10^{53}\frac{\Omega_b}{\Omega_m}
\left(\frac{M}{10^8h^{-1}\,M_\odot}\right)^{5/3}
\left(\frac{\Omega_m}{\Omega_m^z}\frac{\Delta_c}{18\pi^2}\right)^{1/3}
\left(\frac{1+z}{10}\right)h^{-1}\ {\rm erg}.
\label{ebgasz}
\end{equation}
For $z>1$, equations (\ref{tvirz}) and (\ref{ebgasz}) can be simplified
to give equations (\ref{tvir}) and (\ref{ebgas}), respectively, for the 
adopted cosmological parameters.

\clearpage
\figcaption{Schematic diagram of the closed-system model for
(Fe/H) in DLA systems. The baseline (Fe/H)$_P$ is the value
corresponding to a postulated prompt inventory in the IGM.
Normal astration and Fe production by SNe II start in a DLA system
at time $t^*$ after the big bang ($t=0$). Different $t^*$ values for 
different DLA systems result in a wide range of (Fe/H) at a fixed 
$z$. For $t^*=0$, there is a maximum increase in (Fe/H) (trajectory A).}

\figcaption{(a) Evolution of the halo mass $M$ as a function
of $z$ calculated for $1\sigma$ (dot-dashed curve)
and $2\sigma$ (long-dashed curve) halos using the model of
hierarchical structure formation discussed in Barkana \& Loeb (2001).
For a specific halo, this
evolution determines the infall rate. A simple estimate of
the infall rate for $1\sigma$ and $2\sigma$ halos 
can be obtained by taking a typical
slope of $|d\ln M/dz|\sim 2.3$. It is assumed that infall ceases when
$M$ reaches $M_{\rm IC}$. For $M_{\rm IC}=10^{11}\,M_\odot$
(horizontal solid line labeled ``$M_{\rm IC}=10^{11}\,M_\odot$''),
this occurs at redshifts $z_{\rm IC}=1.6$ and 4.3 for $1\sigma$ and 
$2\sigma$ halos, respectively, while for $M_{\rm IC}=10^{12}\,M_\odot$
(horizontal solid line labeled ``$M_{\rm IC}=10^{12}\,M_\odot$''),
this occurs at redshifts $z_{\rm IC}=0.8$ and 2.8 for $1\sigma$ and 
$2\sigma$ halos, respectively. The time at which the virial temperature of 
the gas reaches a critical value $T_{\rm vir,0}$ is taken as the onset 
of astration in a halo. The halo mass $M_0$ required to reach 
$T_{\rm vir,0}$ is shown as a function of $z$ for $T_{\rm vir,0}=300$ K
[solid curve labeled ``$M_0(T_{\rm vir,0}=300$ K)''] and 
$T_{\rm vir,0}=10^4$ K [solid curve labeled 
``$M_0(T_{\rm vir,0}=10^4$ K)''], respectively. The redshift
$z_0$ at which the curve $M_0(z)$ intersects the curve $M(z)$ for a
halo corresponds to the onset of astration. For $T_{\rm vir,0}=300$ K,
$z_0=7.3$ and 16.8 for $1\sigma$ and $2\sigma$ halos, respectively,
while for $T_{\rm vir,0}=10^4$ K, $z_0=3.8$ and 9.7 for $1\sigma$ and 
$2\sigma$ halos, respectively. It is assumed that a single SN II can
expel all the gas from a halo before the binding energy of the gas 
$E_{\rm bi,gas}$ reaches the explosion energy of $10^{51}$ erg for an SN II.
The halo mass $M_{\rm bi}$ required to
reach $E_{\rm bi,gas}=10^{51}$ erg is shown as a function of $z$
(short-dashed curve labeled ``$M_{\rm bi}$''). 
(b) Evolution of [Fe/H] for $1\sigma$
(dot-dashed curve) and $2\sigma$ (long-dashed curve) halos as
functions of $z$. For both halos, it is assumed that the onset of astration 
corresponds to $T_{\rm vir,0}=10^4$ K, instant infall cessation occurs for
$M_{\rm IC}=10^{11}\,M_\odot$, and SNe Ia start to contribute Fe at
[Fe/H]~$=-1$. The solid curve labeled ``no infall'' is for a system
with no infall. For all cases, the Fe production rate per H atom of 
the gas is taken to be $\lambda_{\rm Fe}=(30\ {\rm Gyr})^{-1}$ when
only SNe II provide Fe ([Fe/H]~$<-1$) and $3\lambda_{\rm Fe}$ 
when both SNe II and Ia provide Fe ([Fe/H]~$>-1$).
Using the simple estimate of the infall rate, the case
where the Fe production by SNe II only is balanced by the dilution due to 
infall is shown as the solid curve labeled ``quasi-steady state estimate.''
The overall evolution of [Fe/H] for a $2\sigma$ halo is indicated by 
arrows. Note the sharp initial rise of the evolutionary trajectories for 
both $1\sigma$ and $2\sigma$ halos subsequent to the onset of astration. 
These trajectories
then lie close to the quasi-steady state estimate until infall
cessation ($z_{\rm IC}$). They rapidly approach the case of no infall
after infall cessation ($z<z_{\rm IC}$). (c) Effects of $T_{\rm vir,0}$ and
$M_{\rm IC}$ on the evolution of [Fe/H] for a $2\sigma$ halo.
The solid curve is the case shown in (b). The dot-dashed
curve assumes a different $T_{\rm vir,0}$ of 300 K and the short-dashed 
curve assumes a different $M_{\rm IC}$ of $10^{12}\,M_\odot$. Note that
these three cases share a common period of quasi-steady state evolution.}

\figcaption{(a) Metallicity distribution (in arbitrary units)
of stars in a $2\sigma$ halo. It is assumed that astration starts at
$z_0=9.7$ corresponding to $T_{\rm vir,0}=10^4$ K and infall starts to 
decrease exponentially on a timescale $\tau$ at $z_{\rm IC}=4.3$
corresponding to $M_{\rm IC}=10^{11}\,M_{\odot}$. Only SNe II provide
Fe at a rate of $\lambda_{\rm Fe}=(30\ {\rm Gyr})^{-1}$ per H atom in
the gas. The parameter $\alpha$
for the astration rate is taken to be 1/18.
The sharp peak in the metallicity distribution at [Fe/H]~$=-2.2$
represents the stars formed during the infall phase ($z>z_{\rm IC}$)
while the broad peak at [Fe/H]~$=\log\alpha=-1.3$
represents those formed subsequently ($z<z_{\rm IC}$) until the gas
is exhausted by astration. The rapid drop of the metallicity
distribution at $z=z_{\rm IC}$ is due to the sharp change in
$dZ_{\rm Fe}/dt$ between the regimes with and without infall.
The fraction of stars in the sharp peak is 10\% for the case
of instant infall cessation ($\tau=0$, solid curve). This
fraction is even smaller for finite $\tau$
[dashed ($\tau=0.1$ Gyr) and dot-dashed ($\tau=0.2$ Gyr) curves]
as more stars are formed after infall starts to decrease. 
(b) Solid curve same as (a)
but for $\tau=1$ Gyr. The sharp peak at [Fe/H]~$=-2.2$ is still
present but contains only 2\% of the stars and is subdued as
compared with the cases of rapid infall cessation ($\tau\ll 1$ Gyr). 
The dashed curve assumes a total Fe production rate of $3\lambda_{\rm Fe}$ 
per H atom in the gas for [Fe/H]~$>-1$ due to additional Fe contributions 
from SNe Ia. The sharp drop of the metallicity distribution at [Fe/H]~$=-1$
corresponds to the onset of SNe Ia. Note that a new broad peak is formed
at [Fe/H]~$=\log(3\alpha)=-0.8$. Prior to the onset of SNe Ia 89\% of the 
baryonic matter has been consumed in star formation.
(c) Same as (a) but for a $1\sigma$ halo.
Astration starts at $z_0=3.8$ corresponding to $T_{\rm vir,0}=10^4$ K and
infall starts to decrease exponentially on a timescale $\tau$ 
(solid curve: $\tau=0$, dashed curve: $\tau=1$ Gyr) at 
$z_{\rm IC}=1.6$ corresponding to $M_{\rm IC}=10^{11}\,M_{\odot}$.
Note that the sharp peak corresponding to the time of infall cessation 
is shifted to [Fe/H]~$=-1.8$ as compared to [Fe/H]~$=-2.2$ for the 
$2\sigma$ halo. (d) Same as (a) and (b) but for $z_{\rm IC}=2.8$
corresponding to $M_{\rm IC}=10^{12}\,M_{\odot}$, $\tau=0$, and 
$\alpha=1/6$. Note that the sharp peak corresponding to the time of infall 
cessation is shifted to [Fe/H]~$=-1.9$ as compared to [Fe/H]~$=-2.2$
in (a) and (b) while the broad peak of the solid curve is shifted to
[Fe/H]~$=-0.8$ as compared to [Fe/H]~$=-1.3$ in (a) and (b). Also note
that the new broad peak (dashed curve) due to additional Fe contributions 
from SNe Ia is shifted to [Fe/H]~$=-0.3$ as compared to [Fe/H]~$=-0.8$ in 
(b). Prior to the onset of SNe Ia 45\% of the baryonic matter has been
consumed in star formation.}

\figcaption{Stages of halo evolution. The redshift at which astration
starts in an $n\sigma$ halo is given for $T_{\rm vir,0}=300$ and $10^4$ K, 
respectively, by the correspondingly labeled solid curves. The redshift at 
which an $n\sigma$ halo reaches $E_{\rm bi,gas}=10^{51}$ erg is given by
the correspondingly labeled dot-dashed curve. The redshift at which infall 
ceases for an $n\sigma$ halo is given for $M_{\rm IC}=10^{11}$ and 
$10^{12}\,M_\odot$, respectively, by the correspondingly labeled dashed 
curves. Note that if the onset of astration corresponds to 
$T_{\rm vir,0}=300$ K,
then a single SN II can expel all the gas from a halo, thereby disrupting
its evolution, until $E_{\rm bi,gas}=10^{51}$ erg is reached.
In contrast, if the onset of astration corresponds to 
$T_{\rm vir,0}=10^4$ K, then such disruption cannot occur as
$E_{\rm bi,gas}=10^{51}$ erg is always reached prior to the onset of 
astration. It is expected that $T_{\rm vir,0}$ would increase from 300 to
$10^4$ K as H$_2$ molecules were dissociated during the evolution of
the universe. Thus, the range of halos
that are disrupted by SN II explosions would decrease with time and
eventually vanish as indicated by the curvy arrow.}

\figcaption{Evolution of [Fe/H] as functions of $z$ for $0.67\sigma$
(short-dashed curve), $1\sigma$ (dot-dashed curve), and $2\sigma$ 
(long-dashed curve) halos compared with the data (squares) for 96 DLA 
systems (Prochaska et al. 2003). The evolution of [Fe/H] for all
three halos is calculated using the same assumptions as in Figure 2b
and the dot-dashed, long-dashed, and solid curves are the same as in 
that figure. The prompt inventory proposed earlier (horizontal long-dashed 
line) is shown for reference.}
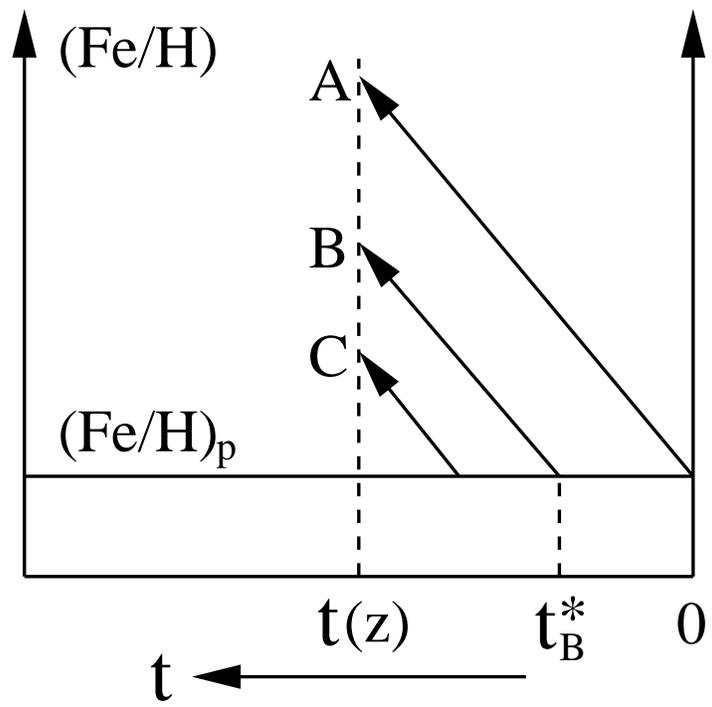
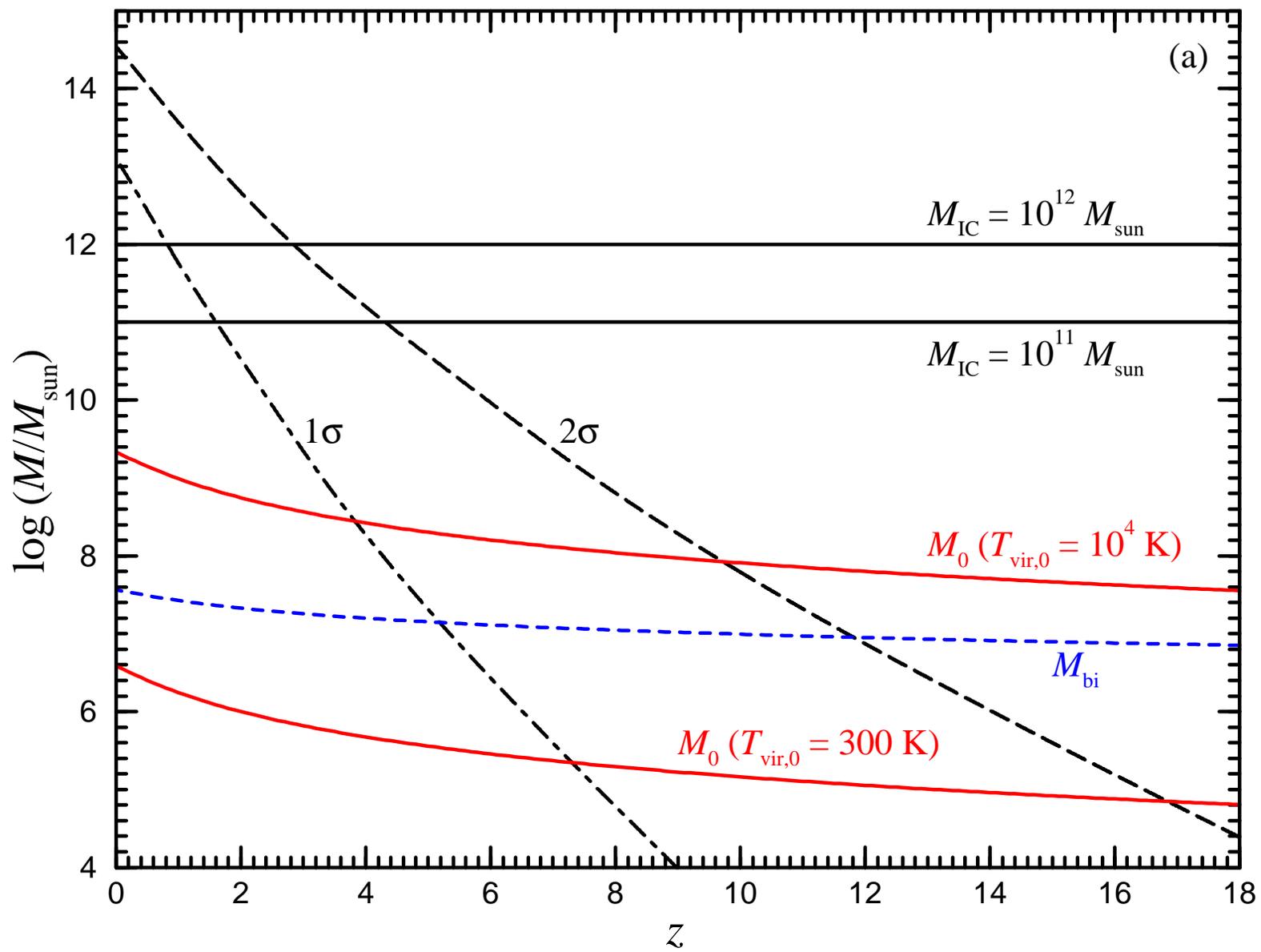
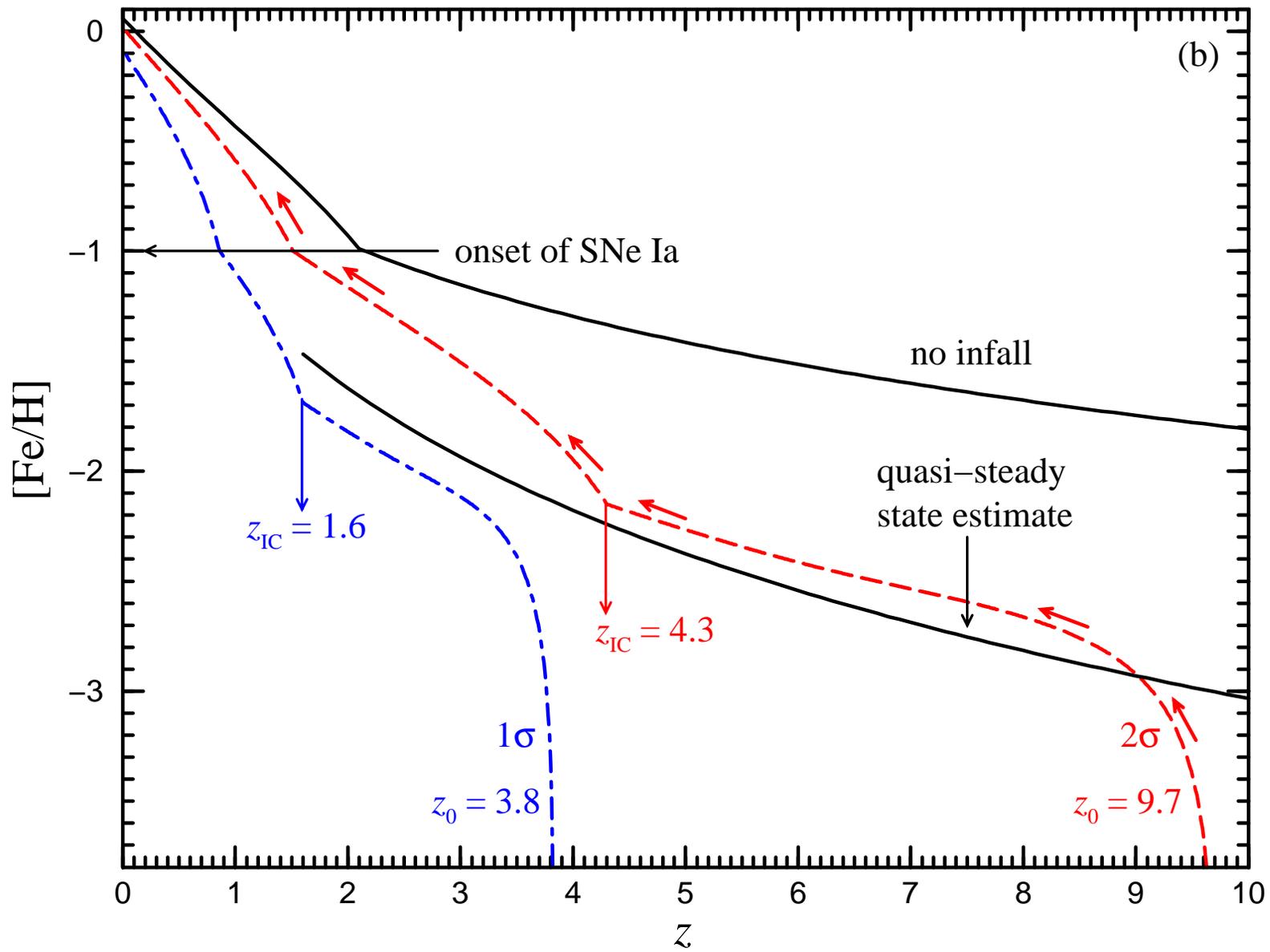
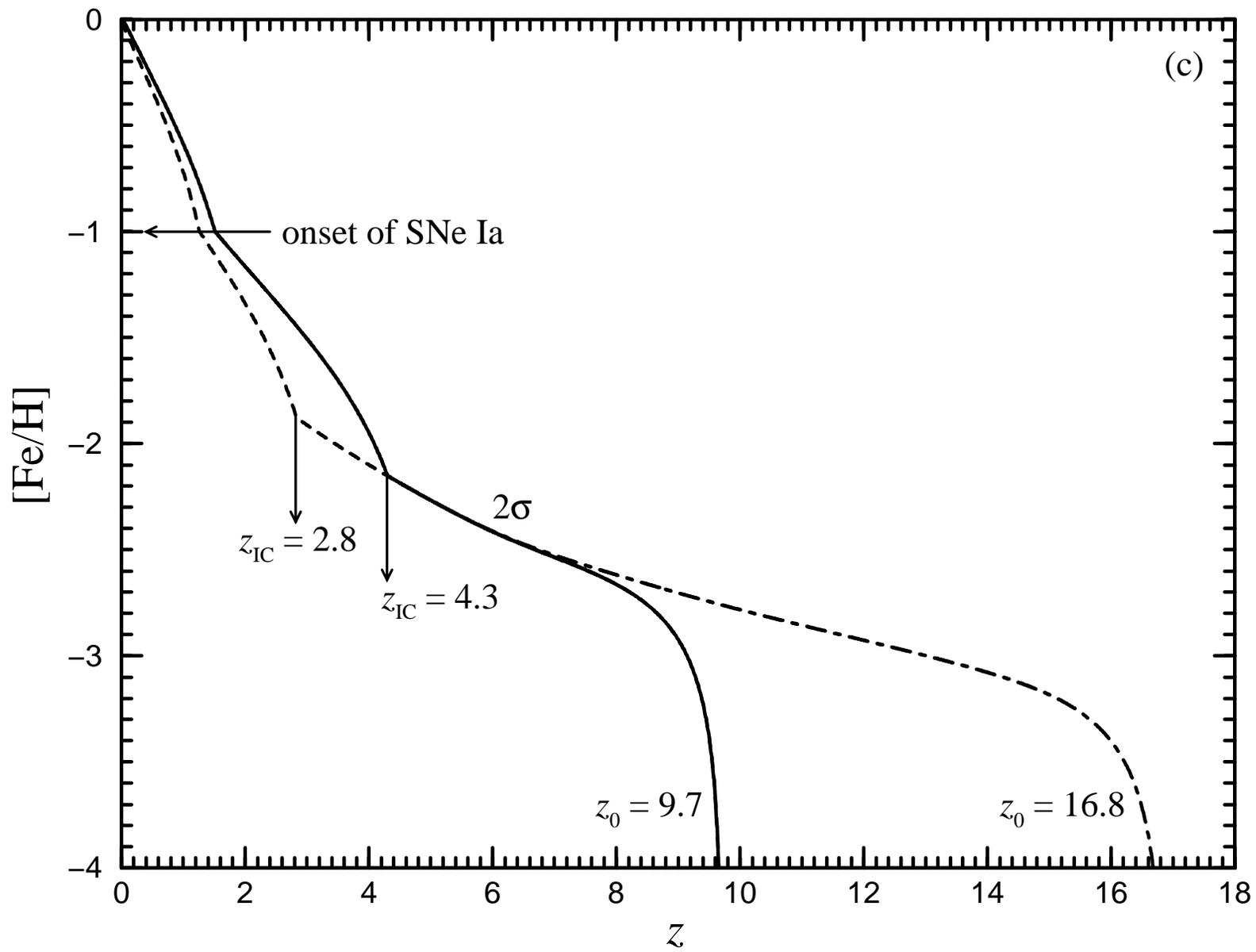
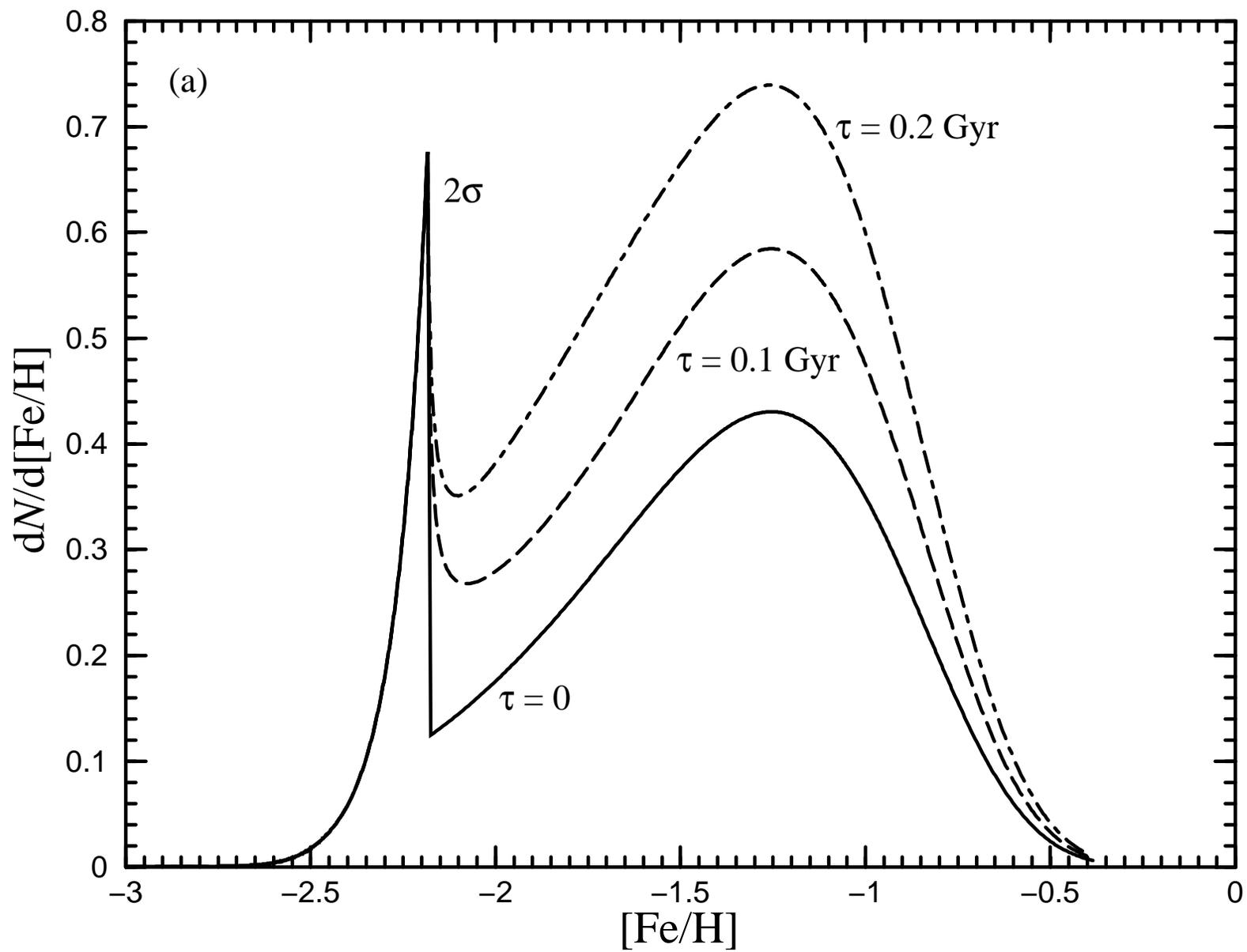
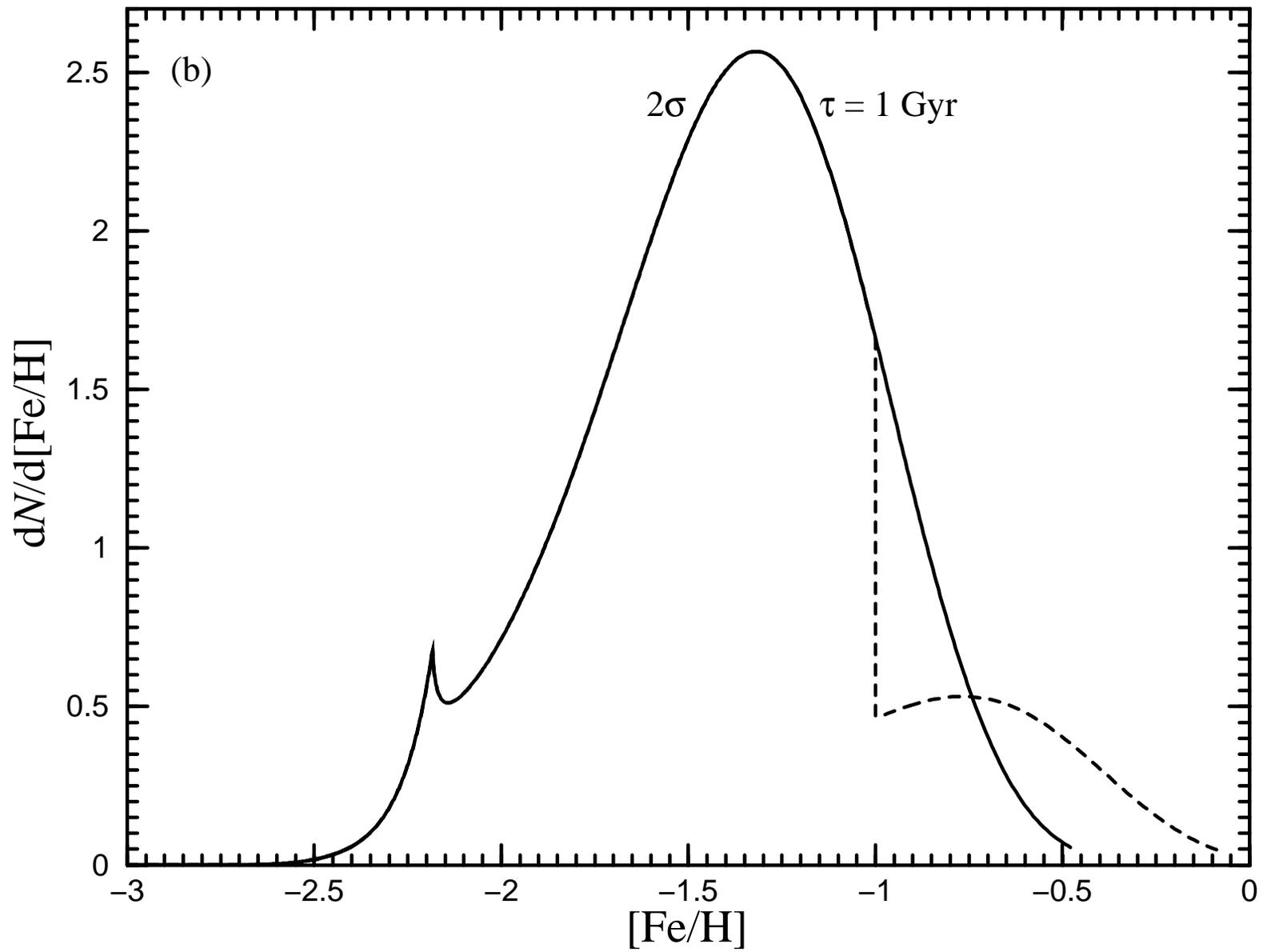
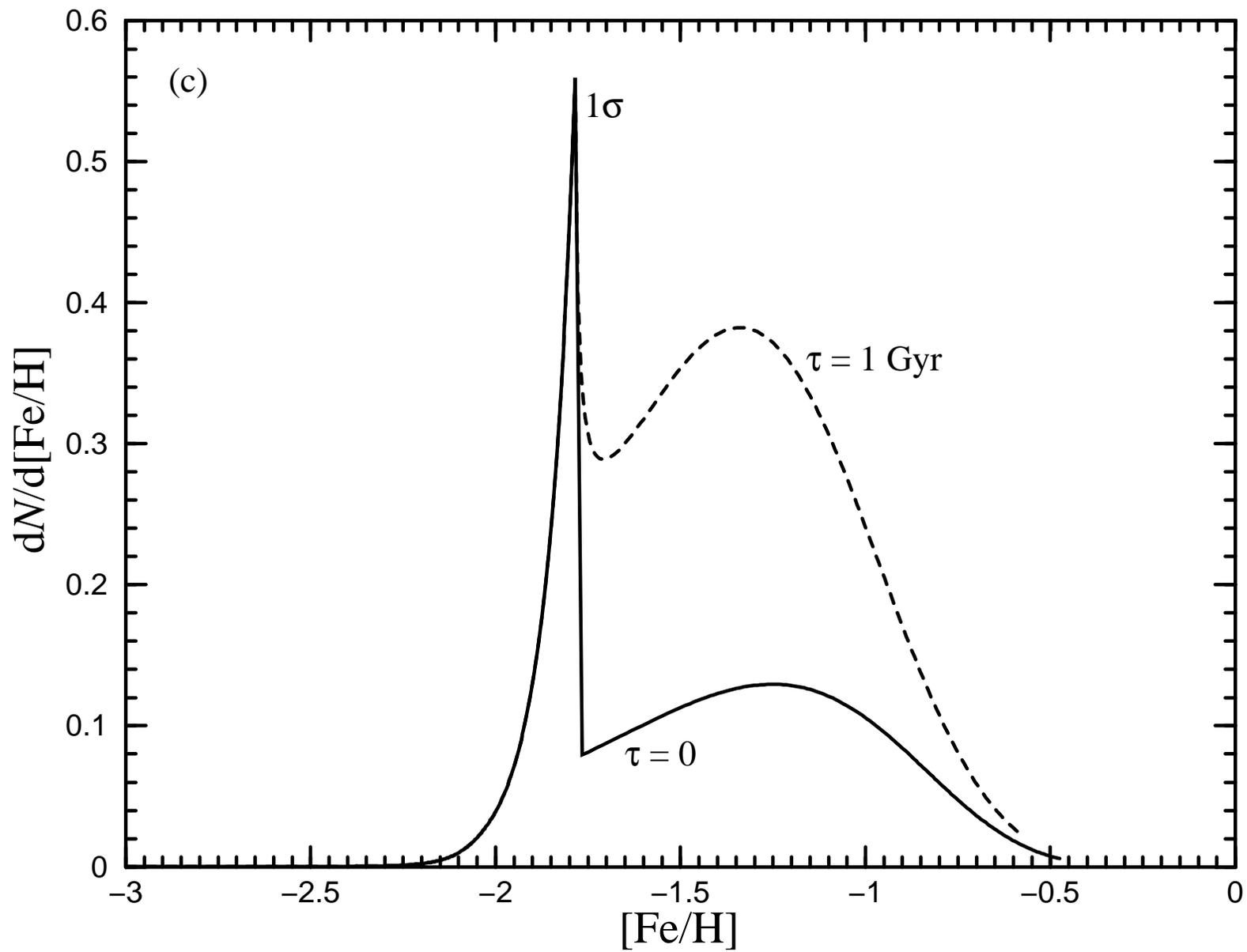
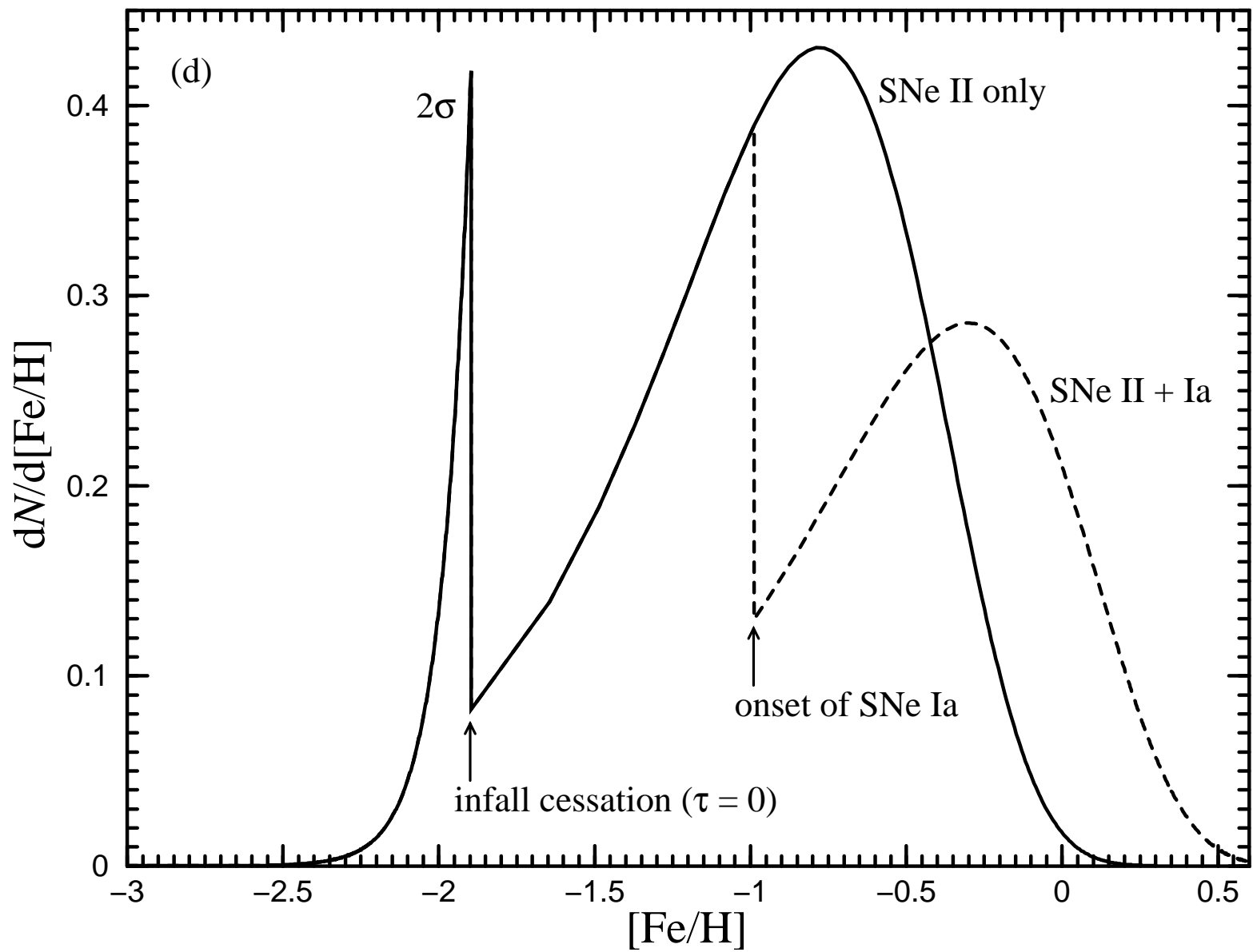
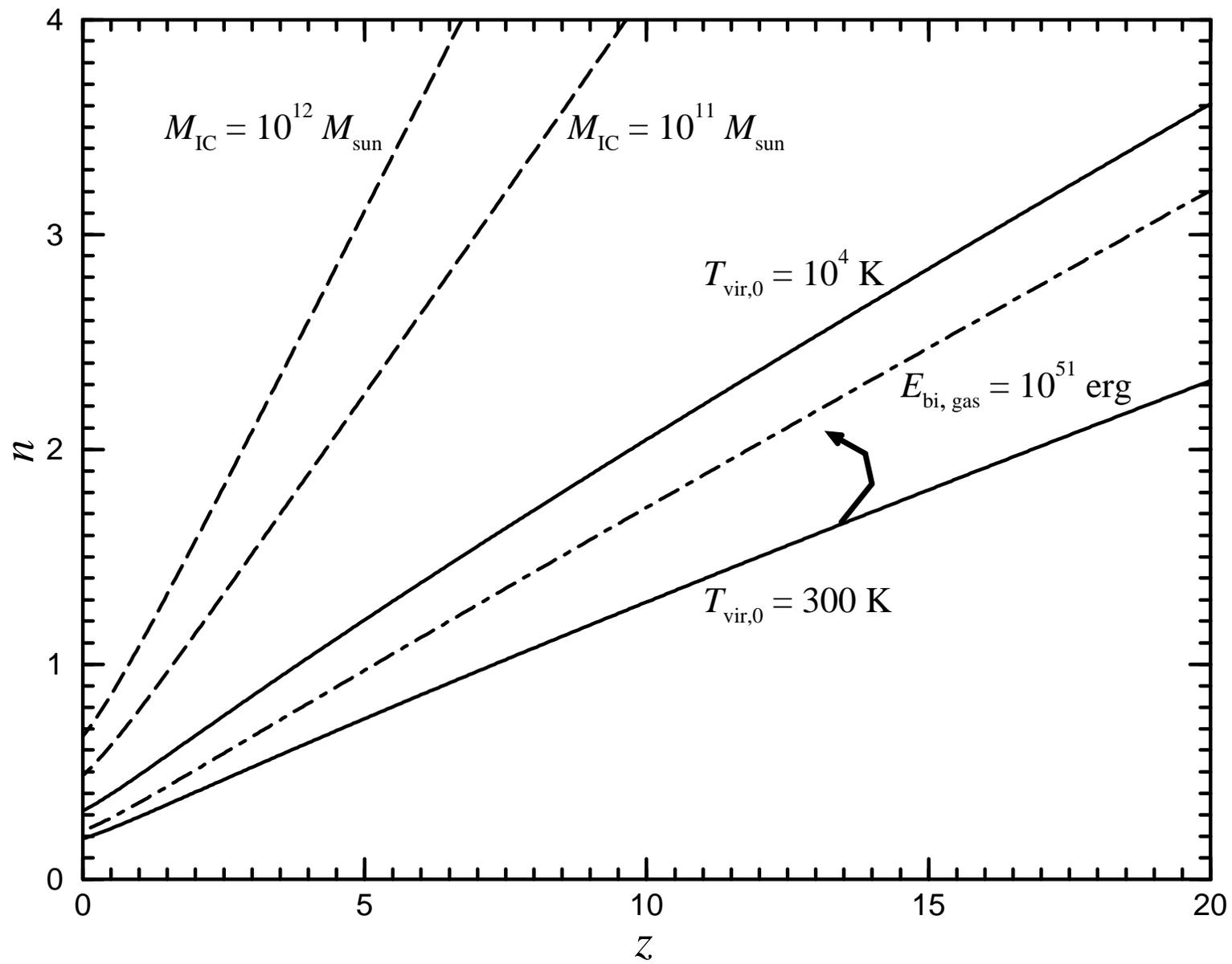
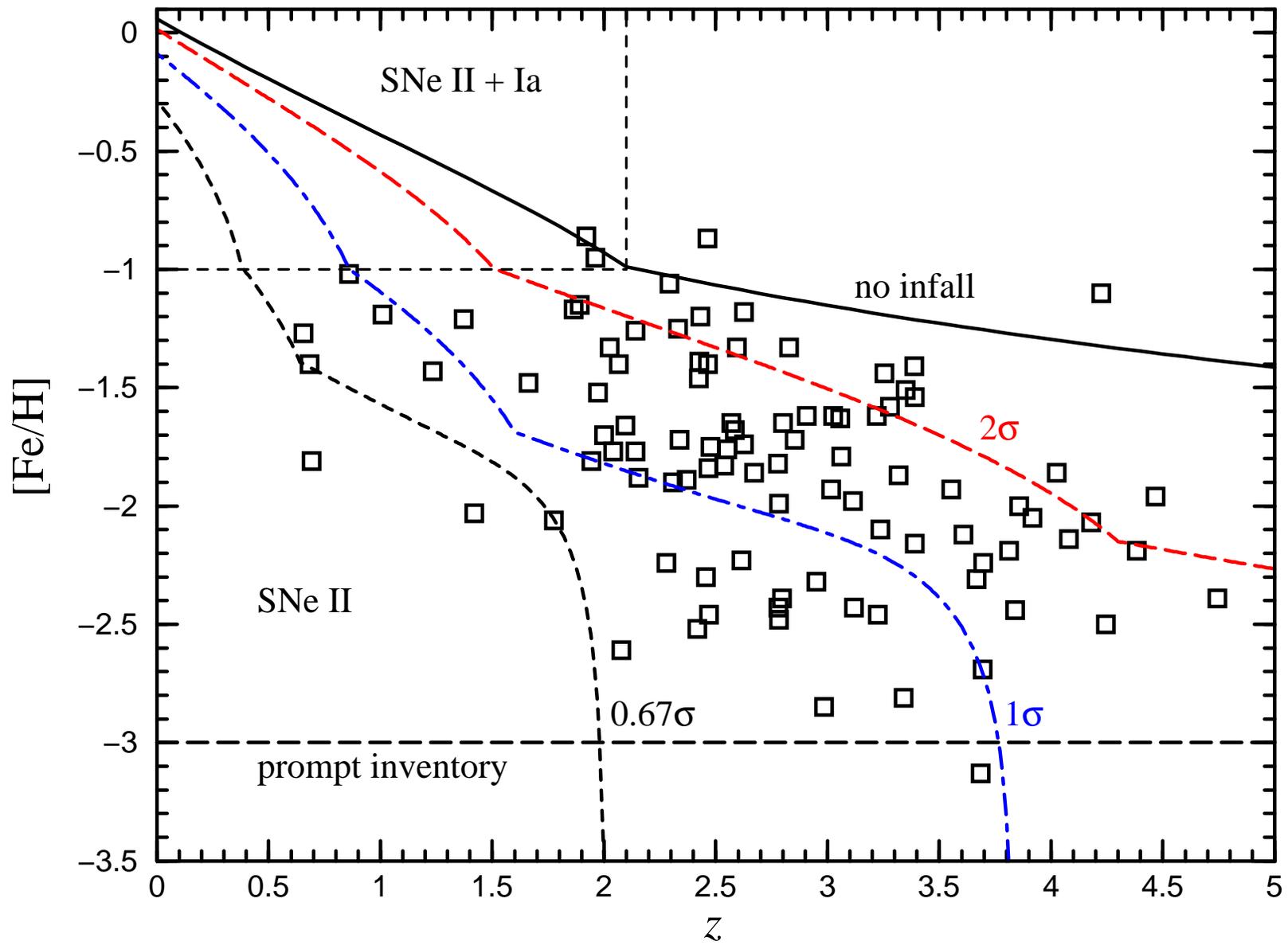
\end{document}